\documentclass{aa}  
\usepackage{amsmath,amssymb}
\usepackage{graphicx}
\usepackage{txfonts}
\usepackage{natbib}
\bibpunct{(}{)}{;}{a}{}{,} 

\usepackage[T1]{fontenc}
\usepackage{txfonts}
\usepackage{textcomp}
\usepackage{gensymb}
\usepackage{verbatim}
\usepackage{lastpage}
\usepackage{fancyhdr}
\usepackage{fancybox}
\usepackage{xspace}
\usepackage{soul}
\usepackage{color}

\RequirePackage{ifpdf}
\ifpdf
  \usepackage[%
    pdftex,%
    colorlinks,%
    hyperindex,%
    plainpages=false,%
    bookmarksopen,%
    bookmarksnumbered%
  ]{hyperref}
  \definecolor{HyperColor}{rgb}{0.0,0.0,0.7}
  \definecolor{CiteColor}{rgb}{0.0,0.0,0.5}
  \hypersetup{%
    linkcolor = {HyperColor},    
    anchorcolor = {HyperColor},  
    citecolor = {CiteColor},     
    filecolor = {HyperColor},    
    menucolor = {HyperColor},    
    pagecolor = {HyperColor},    
    urlcolor = {HyperColor},     
    frenchlinks = false,         
    backref = true,
    pdftitle = {An exosolar systems field-of-view model for exoplanet detection simulations},
    pdfsubject = {},
    pdfkeywords = {Instrumentation: high angular resolution -- Methods: analytical --
astronomical data bases: miscellaneous -- astrometry -- ISM: structure --
Galaxy: stellar content},
    pdfauthor = {A. Belu et al.},
    pdfcreator = {\LaTeX\ with package \flqq hyperref\frqq},
    pdfproducer = {pdfeTeX-0.\the\pdftexversion\pdftexrevision},
  }
\else
  \usepackage{hyperref}
\fi

\newcommand{\D}[1]{\mathrm{d}#1} 
\newcommand{\BF}[1]{\mathbf{#1}} 
\newcommand{\RM}[1]{\mathrm{#1}} 
\newcommand{\dgr}{\hbox{$^\circ$}} 
\newcommand{\micron}{\ensuremath{\micro\mathrm{m}}\xspace}

\newcommand{\Abs}[1]{\left\vert #1\right\vert}

\newcommand{\Lstar}{L_\star}               
\newcommand{\Mstar}{M_\star}               
\newcommand{\Rstar}{R_\star}               
\newcommand{\Tstar}{T_\star}               
\newcommand{\Lsun}{L_\odot}                
\newcommand{\Msun}{M_\odot}                
\newcommand{\Tsun}{T_\odot}                
\newcommand{\Rsun}{R_\odot}                
\newcommand{\ViewDir}{\ensuremath{\mathbf{s}}} 
\newcommand{\RA}{{\ensuremath{RA}}}       
\newcommand{\DEC}{{\ensuremath{DEC}}}     
\newcommand{\BG}{{\mathrm{bg}}}           
\newcommand{\ExoZodi}{{\mathrm{ez}}}      
\newcommand{\LocalZodi}{{\mathrm{z}}}     

\newcommand{\Inorm}{I^\RM{n}_\lambda}        

\newcommand{\Origin}{\textsc{Origin}\xspace}
\newcommand{\Fits}{\textsc{Fits}\xspace}

\definecolor{AlertColor}{rgb}{1,0.60,0.00}
\definecolor{NewColor}{rgb}{0.10,0.60,0.10}
\definecolor{Yellow}{rgb}{0.95,0.95,0.00}

\newcommand{\New}[1]{\textcolor{NewColor}{#1}}

\newcommand{\SIZE}{}
\newcommand{\X}{}
\newcommand{\MC}{}
\newcommand{\EXP}{}

\newcommand{\ROWi}{}
\newcommand{\ROWii}{}

\newenvironment{SmallText}{\fontsize{7}{8.8}}{}

\begin{document}

\title{A scene model of exosolar systems for use in planetary detection and characterisation simulations\thanks{Work supported in part by the ESA/ESTEC
       contract 18701/04/NL/HB, led by Thales Alenia Space.}}

\titlerunning{Exoplanetary \emph{scene} model}

\author{A. Belu \inst{1}
  \and E. Thiébaut \inst{2}
  \and M. Ollivier \inst{3}
  \and G. Lagache \inst{3}
  \and F. Selsis \inst{4}
  \and F. Vakili \inst{1}}

\authorrunning{Belu et al.}

\offprints{A. Belu}

\institute{Laboratoire Universitaire d'Astrophysique (LUAN), Université de
  Nice - Sophia Antipolis and CNRS (UMR 6525), Parc Valrose
  F-06100 Nice, France, \email{belu@unice.fr}
\and Université de Lyon, Lyon, F-69000, France; Université Lyon~1,
  Villeurbanne, F-69622, France; Centre de Recherche Astronomique de Lyon,
  Observatoire de Lyon, 9 avenue Charles André, Saint-Genis Laval cedex,
  F-69561, France; CNRS/UMR-5574; Ecole Normale Supérieure de Lyon, Lyon,
  France
\and Institut d'Astrophysique Spatiale (IAS), bâtiment 121, F-91405 Orsay
  (France), Université Paris-Sud 11 and CNRS (UMR 8617)
\and Centre de Recherche Astrophysique de Lyon (CNRS UMR 5574), Universit 
\'e de Lyon, Ecole Normale Sup\'erieure de Lyon, 46 All\'ee d'Italie  
F-69007 Lyon}

\date{Received ... / accepted ...}


\abstract%
{Instrumental projects that will improve the direct optical finding and characterisation of exoplanets have advanced sufficiently to trigger
 organized investigation and development of corresponding signal processing
 algorithms. The first step is the availability of field-of-view (FOV)
 models. These can then be submitted to various instrumental models, which
 in turn produce simulated data, enabling the testing of processing
 algorithms.}
{We aim to set the specifications of a physical model for typical FOVs of
 these instruments.}
{The dynamic in resolution and flux between the various sources present in such a  FOV imposes a multiscale, independent layer approach. From review of current 
 literature and through extrapolations from currently available data and models, we derive the features of each source-type in the field of view likely to
 pass the instrumental filter at exo-Earth level.}
{Stellar limb darkening is shown to cause bias in leakage calibration if unaccounted
 for. Occurrence of perturbing background stars or galaxies in the typical
 FOV is unlikely. We extract galactic interstellar medium background
 emissions for current target lists. Galactic background can be considered
 uniform over the FOV, and it should show no significant drift with
 parallax. Our model specifications have been embedded into a Java simulator, soon to be made open-source. We have also designed an associated FITS input/output
 format standard that we present here.}
{}


\maketitle


\section{Introduction}

Instruments currently under design for direct optical exoplanetary
search and characterisation need to go beyond the indirect techniques used
so far for the discovery of the $\sim200$ currently known exoplanets, and must
collect planetary photons. Beyond the joint determination of the science objectives of albedo, planetary radius and orbital parameters, the major aim of these instruments is to establish the presence, in a potential atmosphere, of chemical
markers of life processes (biomarkers). This would be done through the
detection of their absorption features in the spectral flux emitted by the
planet.

Because of this, broadband observation is required. Two types of spaceborne instruments are currently under
development. Both types
of instruments reject stellar light, so that the 10$^{-9}$ and
10$^{-6}$ respectively weaker planetary flux is detectable in the residual noise. The \textit{Terrestrial Planet Finder Coronograph} (TPF-C) is a $6\times2~\mathrm{m}^2$ monolithic collector space telescope in the
visible \citep{TraubTPFC_SPIE2006}. Free-flying-collector interferometers,  in the infrared (band extending from 6 to 18 \micron), used in a particular optical design called
nulling interferometry \citep{Bracewell}, are also being considered. In this latter technique, the
optical array is phased so that light from the on-axis star is
destructively interfered. As the array is rotated, the off-axis planets
pass through the peaks and valleys of the instrumental response on the sky
(the so-called \emph{transmission map}), which generates a modulated
signal. The main interferometric projects are \textit{Darwin}
\citep{LegerIcar96,FridlundESA2000} and \textit{TPF-I}
\citep{BeichmanJPLBook99}. Complementarity of biomarkers at these two
wavelength ranges, associated with the advantages and shortcomings of each
of these classes of instruments, explain this parallel effort. 

Both approaches are currently mature enough to trigger organized
investigation and development of signal processing algorithms for planet
detection and characterisation: \citet{FerrariUAI} for direct imaging, and
\citet{MugnierITHD}, \citet{Thiebaut}, \citet{Thiebaut2007}, \citet{Marsh}, \citet{Draper} for nulling interferometry.

This paper specifies a physical and mathematical model of
source FOVs, called \Origin.  As will be seen in the various sections of
this paper, there is  an abundance of available elements
characterizing exoplanetary FOVs. We felt there was a need  for an
integrated access to this information for simulation input, data exchange
and outreach.

At the present time, the instruments capable of exo-Earth detection and characterisation are in
a very early definition phase: no concepts are considered final. For this reason we make no simplifying assumption regarding the instrument, in
particular its sensitivity and/or its ability to discriminate between specific scene features and/or noise sources.

This work has been greatly inspired by the European \textit{Darwin}
mission, hence the nulling interferometry point of view is often significantly
developed beyond the conclusions that apply more generally to exo-Earth finding
instruments.

\section{Framework}
\label{s:frame}

In this section we present the framework elements of our model: the
building-block rationale and considerations on spatial, spectral
and temporal resolutions.

\subsection{Building-Block Model}
\label{s:building-block}

Our simulator of astronomical \emph{scenes} aims at modeling the
angular and spectral distribution of light received from the observed
exoplanetary system.  The proposed model is built by superimposing the
emission of the various sources that are seen by the instrument.
Following this, the specific intensity (units:
$\mathrm{W\,m^{-2}\,rad^{-2}\,\micron^{-1}}$) observed in a direction
$\ViewDir$ is:
\begin{eqnarray}
  I(\ViewDir,\lambda,t) &=& I_\star(\ViewDir,\lambda,t)
  + \sum_{j} I_j(\ViewDir,\lambda,t) \nonumber \\
  & & \mbox{} + I_\ExoZodi(\ViewDir,\lambda,t)
  + I_\LocalZodi(\lambda,t)
  + I_\BG(\ViewDir,\lambda,t)\,,
  \label{e:frame}
\end{eqnarray}
where $I_\star$ is the star's emission (see Sect.~\ref{s:star}), $I_j$ is the
contribution of the $j$-th planet (see Sect.~\ref{s:expn}), $I_\ExoZodi$ is
the emission by the exozodiacal cloud (see Sect.~\ref{s:exo-zodi}),
$I_\LocalZodi$ is the local zodiacal cloud emission assumed to be uniform
across the field of view (see Sect.~\ref{s:local_zodi}) and, finally, $I_\BG$
accounts for the contribution of background sources (Sect.~\ref{s:ISM} and Sect.~\ref{s:bkgrd_obj}). If $(x,y)$ are the cartesian coordinates of a given
source in a plane perpendicular to the line of sight, then the angular direction of observation with respect to the center of the field of view is $\ViewDir\simeq\New{(\theta_\RM{x},\theta_\RM{y})=}\,(x,y)/d$, where $d$ is the distance from the observer.

For each source type, the various sections of this paper will present
a discussion on the astrophysical features likely to pass the instrumental
filter, at the level of the signal that an exo-Earth would produce, and hence
which require modeling; subsequently, example layer-outputs for that particular
source are demonstrated.

\subsection{Coordinates Systems}
\label{s:coordsyst}

Our model accounts for the observatory's Solar-system
coordinates (ecliptic, Earth or L2\footnote{Second Lagrange point of the
Earth-Sun 2-bodies system}) and for target coordinates, enabling local zodiacal drift
computing (see Sect.~\ref{s:local_zodi}). The planed vertical extent (under $7 \times 10^{5}$ km ) of the observatory's halo Lissajous libration orbit around L2 \citep{landgraf_orbit} is considered small compared to the zodiacal cloud's thickness, and was not implemented.

A target (\textit{i.e.}  exosystem)-proper coordinate system is implemented
for calculation of the stellar flux received and phase-reflected by a
planet (Sect.~\ref{s:rflctd}). It also enables consistent scene generation
for revisits, astrophysical community databases import/export, and robustness testing of image reconstruction algorithms.

Temporal accounting has two aspects. \textit{Time} is used to model
variations in the scene that occur typically on timescales comparable to
those of an observation, such as the motion of close, short revolution-period
planets (Pegasides), or stellar variability (flares). The resolution is typically
1/10$^{\mathrm{th}}$ of an hour. \textit{Dates} are used to initialize scene
consistently, for simulating revisits; the same 1/10$^{\mathrm{th}}$ hour resolution applies, but expressed in fractional Julian Date (JD).  In the output of our model, we assume that the exposure duration $\Delta{}t$ is sufficiently short to
consider that the scene is static during a given exposure.

\subsection{Spatial Resolution}
\label{s:spatial-resolution}

In the output of the proposed building-block model (see
Appendix~\ref{s:output}), sources with apparent size larger than the
instrumental resolution must be considered as \emph{resolved}, and
modeled by maps of their brightness distribution.  These maps (or
\emph{images}) must be sampled with a pixel size $\Delta\theta$ smaller
than the instrumental resolution limit:
\begin{equation}
  \Delta\theta \ll \frac{\lambda_{\RM{min}}}{B_{\RM{max}}}
\end{equation}
where $\lambda_{\RM{min}}$ is the smallest spectral channel effective wavelength, and $B_{\RM{max}}$
the largest interferometric baseline.  For a monolithic telescope, 
$B_{\RM{max}}$ is the diameter of the pupil.

\subsection{Spectral Resolution}
\label{s:spectral-resolution}

To avoid loss of coherence in the interferometer model due to the finite
spectral bandwidth $\Delta\lambda$ in the scene model, the spectral
resolution must be chosen so that the phase difference across a spectral
channel is negligible:
\begin{equation}
  \Abs{\frac{B\,\theta}{\lambda - \frac{1}{2}\,\Delta\lambda} -
       \frac{B\,\theta}{\lambda + \frac{1}{2}\,\Delta\lambda}}
  \simeq \frac{B\,\theta\,\Delta\lambda}{\lambda^2}
  \ll 1
\end{equation}
where $\lambda$ is the  effective wavelength of the spectral channel, $B$ is the interferometric baseline, and $\theta$ is the view angle.
Hence the spectral resolution must be chosen so that:
\begin{equation}
  \frac{\lambda}{\Delta\lambda} \gg
  \frac{B_{\mathrm{max}}\,\theta_{\mathrm{max}}}{\lambda}
\end{equation}
where $\theta_{\mathrm{max}}$ is the radius of the field of view.

The interferometric baseline is typically chosen so as to have the dark
fringe of the nulling interferometer not larger than the inner size
$\theta_{\RM{HZ}_\RM{in}}$ of the \emph{habitable zone}, that is:
$\lambda/B_{\RM{max}}\sim\theta_{\RM{HZ}_\RM{in}}$.  Assuming a typical
field of view radius of $\theta_{\mathrm{max}}\sim0\farcs5$, and since the
smallest considered $\theta_{\RM{HZ}_\RM{in}}$ is $0\farcs01~mas$
\citep{Kaltenegger}, the spectral resolution must be better than $\sim50$.

\subsection{Photon Counts}
\label{s:photon-counts}

To produce an image, it is convenient to first consider the expression of the photon count received into a spectral channel by a pixel during an exposure. For a \emph{resolved} source this is:

\begin{equation}
  N_\RM{resolved} = \eta_\lambda \, \mathcal{S}_\RM{tel} \,
  \frac{\lambda}{h\,c} \,
  I(\ViewDir,\lambda,t) \, \Delta\theta^2 \, \Delta\lambda \, \Delta{}t\,,
  \label{eq:resolved-count-per-pixel}
\end{equation}
where $\Delta{}t$ is the exposure duration,
$\mathcal{S}_\RM{tel}$ is the collecting area (\textit{e.g.}\ unobstructed
surface of the telescope primary mirror) and $\eta_\lambda$ is the
instrumental throughput at the spectral channel wavelength.  The photon
count in Eq.~(\ref{eq:resolved-count-per-pixel}) assumes that the spectral
bandwidth $\Delta\lambda$, the pixel size $\Delta\theta$, and the exposure
$\Delta{}t$ are small enough compared to typical scales of variation of the
specific intensity $I(\ViewDir,\lambda,t)$, as discussed in the previous
subsections.

For an \emph{unresolved} source (\textit{e.g.}\ a planet or any point-like
source), the term $I(\ViewDir,\lambda,t)\,\Delta\theta^2$ in
Eq.~(\ref{eq:resolved-count-per-pixel}) must be replaced by the specific
flux $F(\lambda,t)$. For the planets, the specific flux can be computed
straightforwardly from atmospheric models (Sect.~\ref{s:franck}) which are
used as input databases in our building-block model. The number of photons
received by a pixel is thus:

\begin{equation}
  N_\RM{unresolved} = \eta_\lambda \,
  \mathcal{S}_\RM{tel} \, \frac{\lambda}{h\,c} \, F(\lambda,t) \,
  \Delta\lambda \, \Delta{}t \,.
  \label{eq:unresolved-count-per-pixel}
\end{equation}

The building-block output of our model comprises two
categories, accordingly. For \emph{resolved} sources such as the exozodiacal dust
emission, the output consists of 3-D
$(\theta_\RM{x},\theta_\RM{y},\lambda)$ maps of the brightness distribution
of the source at a given observing time $t$, and integrated by a pixel of
angular size $\Delta\theta$ and effective spectral bandwidth
$\Delta\lambda$:
\begin{equation}
  f_\RM{resolved}(\ViewDir,\lambda,t) =
  \frac{\lambda\,I(\ViewDir,\lambda,t)}{h\,c}\,\Delta\theta^2\,\Delta\lambda
  \label{eq:resolved-output}
\end{equation}
in units of number of incident photons per $\mathrm{m}^2$ per $\mathrm{s}$.
For \emph{unresolved} sources, such as the planets, the building-block
model output is simply the specific flux of the point-like source
integrated in every spectral channel at a given observing time $t$:
\begin{equation}
  f_\RM{unresolved}(\lambda,t) =
  \frac{\lambda\,F(\lambda,t)}{h\,c}\,\Delta\lambda
  \label{eq:unresolved-output}
\end{equation}
in the same units as $f_\RM{resolved}$.  

\section{Star Model}
\label{s:star}

This section describes the modeling of $I_\star(\ViewDir,\lambda,t)$, the
specific intensity emitted by the star.

\subsection{Limb Darkening and Stellar Leakage}
\label{s:limb-darkening}

Due to the finite extension of the star, coupled with instrumental instabilities, the stellar
light may not be completely suppressed by the coronagraphic technique, be it classical or interferometric coronagraphy. The
residual light is called leakage. We would like to know if choosing not to model the stellar limb darkening would lead to significant errors of the leakage that an instrumental simulator would produce.

Assuming the star has spherical symmetry, the angular and spectral
distribution of its emission is given by:
\begin{equation}
  I_{\star}(\mu,\lambda) = I_{\lambda}^\RM{n}\,D_{\lambda}(\mu)
\end{equation}
where $I_{\lambda}^\RM{n}=I_{\star}(\mu=0,\lambda)$ is the specific
intensity emitted by the star in a direction normal to its surface, and
$D_{\lambda}(\mu)$ is the limb-darkening law which is a function of the
cosine $\mu$ of the angle between the viewing direction and the normal to
the surface:
\begin{equation}
  \mu = \sqrt{1-(\theta/\theta_\star)^2}
\end{equation}
where $\theta$ is the angular direction with respect to the center of the
star and $\theta_\star$ is the apparent radius of the star.  Following
\citet{VanHamme}, we consider the following possible limb-darkening laws:
\begin{equation}
  D_{\lambda}\left(\mu\right) = 
  \left\{\begin{array}{ll}
    1 
    & \mbox{\small(black body)} \\
    1 - x_\lambda\,(1 - \mu) 
    & \mbox{\small(linear)} \\
    1 - x_\lambda\,(1 - \mu) - y_\lambda\,\mu\,\log\mu
    & \mbox{\small(logarithmic)} \\
    1 - x_\lambda\,(1 - \mu) - y_\lambda\,(1 - \sqrt{\mu}) 
    & \mbox{\small(square root)} \\    
  \end{array}\right.
\end{equation}
where $x_\lambda$ and $y_\lambda$ are tabulated parameters which have been
computed by \citet{VanHamme} for 410 stellar spectra synthesized with the
ATLAS code.

Now we consider the case of a nulling interferometer. The stellar
contribution in its output signal reads:
\begin{equation}
  \label{eq:star-leakage}
  A_\star(\lambda,t)
  = \int R(\ViewDir,\lambda,t)\,I_\star(\ViewDir,\lambda,t)
  \,\D{\ViewDir}
\end{equation}
where $R(\ViewDir,\lambda,t)$ is the instrumental instantaneous  
transmission map. At the center of the field of view, it can be
approximated \citep{Absil-2001-thesis-nulling_interferometry}, in the ideal, unperturbed case, by a power
law. We also consider an azimuthal symmetry (similar results are obtained in a full two-dimensional integration) which leads to:
\begin{equation}
  R(\ViewDir,\lambda,t) \propto\theta^{n}\,.
\end{equation}

We are now going to take this pattern as a model for the effective transmission over a realistic exposure, with perturbations. \citet[Table 5]{lay_2004} shows that, in total, power-law "geometric type" leakage contributions are dominant over the "floor type" contributions. Indeed, we are not interested in the exact evaluation of the leakage, but in a useful approximation of the \emph{relative} influence of modeling the integration of the limb-darkened emission through a "null profile".

Then, since $\theta=\theta_\star\,\sqrt{1-\mu^2}$ and
assuming, for sake of simplicity, a linear limb-darkening law, the star
leakage scales as:
\begin{eqnarray}
  A_\star(\lambda,t)
  &\propto& \int \left(1 - \mu^2\right)^{n/2}\,
  \left[1 - (1 - \mu)\,x_{\lambda}\right]\,\mu\,\D{\mu}\nonumber\\
  &=&\frac{1 - \gamma(n)\,x_{\lambda}}{n + 2}
  \label{eq:linear-leakage}
\end{eqnarray}
with
\begin{equation}
  \label{e:gamma}
  \gamma(n) = 1 - \frac{n+2}{2}\,
  \frac{
    \Gamma\left(\frac{3}{2}\right)\,
    \Gamma\left(\frac{n + 2}{2}\right)
  }{
    \Gamma\left(\frac{n + 5}{2}\right)
  }
\end{equation}
and where $\Gamma$ is the gamma function:
\begin{equation}
  \Gamma(z) = 
  \int_{0}^{+\infty}t^{z-1}\,\mathrm{e}^{-t}\,\mathrm{d}t\,.
  \label{eq:Gamma-law}
\end{equation}

\begin{table}[ht]
  \caption{Numerical values of $\gamma(n)$ defined in Eq.~(\ref{e:gamma})
    for different nulling power $n$.}
  \label{t:gamma}
  \centering
  \begin{tabular}{r@{~:~}ccc}
    \hline
    $n$ & 2 & 4 & 6 \\
    $\gamma(n)$ & 0.467 & 0.543 & 0.594 \\
    \hline
  \end{tabular}
\end{table}

To assume that the star emits as a black body is equivalent to taking
$x_\lambda=0$. Thus the term $\gamma(n)\,x_{\lambda}$ in
Eq.~(\ref{eq:linear-leakage}) is equal to the relative attenuation of
stellar leakage due to the limb-darkening law, compared to a pure black
body star.  Table~\ref{t:gamma} shows that $\gamma(n)\simeq0.5$ for typical
values of $n$, and Table~\ref{t:mean-xlin} lists the average value of
$x_\lambda$ in the photometric bands.  From these tables, we can estimate
that the \emph{relative} attenuation of stellar leakage due to taking into account the limb-darkening
law is $\gamma(n)\,x_{\lambda}\sim5\,\%$ in the infrared for F to M stars,
but can be as high as $\sim35\,\%$ in the visible. These figures are consistent with similar analysis by \citet{Absil_ground_AA}. Although computed for a
linear law, the same results would be obtained for the other limb-darkening laws considered and, to summarize, accounting for the limb darkening is
important so as not to overestimate the stellar leakage in the visible range, but
could be neglected, in a first approximation, in the infrared.  Finally it
is worth noting that dynamic errors over the duration of an exposure
will result in more stellar leakage, some of which can not be easily
calibrated \citep{lay_2004}.

\begin{table*}
  \renewcommand{\SIZE}[1]{{\tiny #1}}
  \renewcommand{\MC}[3]{\multicolumn{#1}{#2}{\SIZE{#3}}}
  \renewcommand{\X}[1]{& #1}
  \caption{Mean linear limb-darkening coefficient $x_\lambda$ in
    photometric bands for stars of different spectral types.}
  \label{t:mean-xlin}
  \centering
  \begin{tabular}{rllllllllllll}
    \hline\hline
    \MC{1}{c}{Type} &
    \MC{1}{c}{U} &
    \MC{1}{c}{B} &
    \MC{1}{c}{V} &
    \MC{1}{c}{R} &
    \MC{1}{c}{I} &
    \MC{1}{c}{J} &
    \MC{1}{c}{H} &
    \MC{1}{c}{K} &
    \MC{1}{c}{L} &
    \MC{1}{c}{M} &
    \MC{1}{c}{N} &
    \MC{1}{c}{Q} \\
    \hline
    \SIZE{F0V} \X{0.55} \X{0.57} \X{0.48} \X{0.37} \X{0.29} \X{0.23}
               \X{0.17} \X{0.15} \X{0.13} \X{0.11} \X{0.06} \X{0.04}\\
    \SIZE{F2V} \X{0.58} \X{0.59} \X{0.49} \X{0.38} \X{0.30} \X{0.24}
               \X{0.18} \X{0.16} \X{0.13} \X{0.11} \X{0.06} \X{0.04}\\
    \SIZE{F5V} \X{0.67} \X{0.64} \X{0.52} \X{0.42} \X{0.33} \X{0.27}
               \X{0.20} \X{0.17} \X{0.14} \X{0.12} \X{0.06} \X{0.04}\\
    \SIZE{F8V} \X{0.72} \X{0.67} \X{0.55} \X{0.44} \X{0.35} \X{0.28}
               \X{0.21} \X{0.18} \X{0.15} \X{0.12} \X{0.07} \X{0.05}\\
    \SIZE{G0V} \X{0.78} \X{0.71} \X{0.57} \X{0.46} \X{0.37} \X{0.30}
               \X{0.22} \X{0.19} \X{0.16} \X{0.13} \X{0.07} \X{0.05}\\
    \SIZE{G2V} \X{0.84} \X{0.75} \X{0.61} \X{0.49} \X{0.40} \X{0.32}
               \X{0.24} \X{0.21} \X{0.17} \X{0.13} \X{0.07} \X{0.05}\\
    \SIZE{G5V} \X{0.84} \X{0.75} \X{0.61} \X{0.49} \X{0.40} \X{0.32}
               \X{0.24} \X{0.21} \X{0.17} \X{0.13} \X{0.07} \X{0.05}\\
    \SIZE{G8V} \X{0.89} \X{0.79} \X{0.64} \X{0.52} \X{0.42} \X{0.34}
               \X{0.25} \X{0.22} \X{0.17} \X{0.14} \X{0.08} \X{0.05}\\
    \SIZE{K0V} \X{0.94} \X{0.83} \X{0.68} \X{0.55} \X{0.45} \X{0.36}
               \X{0.27} \X{0.23} \X{0.18} \X{0.14} \X{0.08} \X{0.05}\\
    \SIZE{K1V} \X{0.97} \X{0.87} \X{0.73} \X{0.58} \X{0.47} \X{0.38}
               \X{0.28} \X{0.25} \X{0.19} \X{0.14} \X{0.08} \X{0.05}\\
    \SIZE{K2V} \X{0.97} \X{0.87} \X{0.73} \X{0.58} \X{0.47} \X{0.38}
               \X{0.28} \X{0.25} \X{0.19} \X{0.14} \X{0.08} \X{0.05}\\
    \SIZE{K3V} \X{1.00} \X{0.91} \X{0.76} \X{0.62} \X{0.49} \X{0.41}
               \X{0.30} \X{0.26} \X{0.20} \X{0.15} \X{0.09} \X{0.05}\\
    \SIZE{K4V} \X{1.01} \X{0.95} \X{0.80} \X{0.64} \X{0.51} \X{0.43}
               \X{0.31} \X{0.27} \X{0.21} \X{0.16} \X{0.09} \X{0.05}\\
    \SIZE{K5V} \X{0.98} \X{0.94} \X{0.79} \X{0.66} \X{0.51} \X{0.43}
               \X{0.32} \X{0.27} \X{0.21} \X{0.16} \X{0.09} \X{0.06}\\
    \SIZE{K7V} \X{0.85} \X{0.82} \X{0.70} \X{0.61} \X{0.46} \X{0.37}
               \X{0.31} \X{0.25} \X{0.18} \X{0.14} \X{0.09} \X{0.06}\\
    \SIZE{M0V} \X{0.73} \X{0.68} \X{0.61} \X{0.56} \X{0.40} \X{0.29}
               \X{0.24} \X{0.19} \X{0.14} \X{0.13} \X{0.08} \X{0.06}\\
    \SIZE{M1V} \X{0.73} \X{0.68} \X{0.61} \X{0.56} \X{0.40} \X{0.29}
               \X{0.24} \X{0.19} \X{0.14} \X{0.13} \X{0.08} \X{0.06}\\
    \SIZE{M2V} \X{0.73} \X{0.66} \X{0.62} \X{0.57} \X{0.40} \X{0.25}
               \X{0.18} \X{0.15} \X{0.12} \X{0.12} \X{0.07} \X{0.06}\\
    \SIZE{M3V} \X{0.73} \X{0.66} \X{0.62} \X{0.57} \X{0.40} \X{0.25}
               \X{0.18} \X{0.15} \X{0.12} \X{0.12} \X{0.07} \X{0.06}\\
    \SIZE{M4V} \X{0.72} \X{0.63} \X{0.58} \X{0.54} \X{0.37} \X{0.23}
               \X{0.17} \X{0.14} \X{0.11} \X{0.11} \X{0.07} \X{0.06}\\
    \hline
  \end{tabular}
\end{table*}

\subsection{Spectral Classification}
\label{s:spectral-classification}

In the implementation of our scene model, we make use of an input database
of stellar parameters, for different stellar spectral types.  This database
is derived from the work of \citet{VanHamme}, who computed the detailed
coefficients for various limb-darkening laws of synthetic spectra from the
ATLAS code. These spectra were simulated for solar chemical composition
stars with a wide range of effective temperatures $T_\star$, and surface
gravities $g_\star$, covering most of the observed HR diagram.

By using spectral classification tables \citep{SK}, we derived the MK
spectral type and luminosity class of the stars from their physical
parameters $g_\star$ and $T_\star$.  The classification also yields the
stars' absolute luminosity $L_\star$, and an estimate of their mass $M_\star$
which is required to compute the orbits of the planets.  Note that the MK
classification only gives average or typical values of physical stellar
parameters for a given spectral type; these values are therefore not
consistent with those of an individual star.  To reduce these
inconsistencies, we tuned the parameters so that the star radius and
surface gravity verify:
\begin{equation}
  \label{e:Rstar}
  \Rstar = \Rsun\,\left(\frac{\Lstar}{\Lsun}\right)^{1/2}\,
  \left(\frac{\Tsun}{\Tstar}\right)^2\,,
\end{equation}
and
\begin{equation}
  \label{e:Gstar}
  g_\star = g_{\sun}\,\frac{\Mstar}{\Msun}\,
  \left(\frac{\Rsun}{\Rstar}\right)^2\,.
\end{equation}
Table~\ref{t:star-params} lists the resulting star parameters in our input
database for spectral types F through M.

\begin{table}
  \renewcommand{\SIZE}[1]{{\tiny #1}}
  \renewcommand{\MC}[3]{\multicolumn{#1}{#2}{\SIZE{#3}}}
  \renewcommand{\EXP}[1]{\cdot10^{#1}}
  \renewcommand{\ROWi}[6]{\SIZE{#1} & \SIZE{$#2$} & \SIZE{$#3$} & \SIZE{$#4$} & \SIZE{$#5$} & \SIZE{$#6$}\\}
  \renewcommand{\ROWii}[1]{}
  \caption{Star model parameters in the \Origin database for spectral types F
    through M.}
  \label{t:star-params}
  \centering
  \begin{tabular}{rrlllr}
    \hline\hline
    \MC{1}{c}{Type} &
    \MC{1}{c}{$T_\star$} &
    \MC{1}{c}{$M_\star/M_\odot$} &
    \MC{1}{c}{$L_\star/L_\odot$} &
    \MC{1}{c}{$R_\star/R_\odot$} &
    \MC{1}{c}{$\log(g_\star/g_\odot)$} \\
    \hline
    \ROWi{F0V}{7250}{1.60}{5.61}{1.50}{-0.15}
    \ROWi{F2V}{7000}{1.52}{4.35}{1.42}{-0.12}
    \ROWi{F5V}{6500}{1.40}{2.72}{1.30}{-0.08}
    \ROWi{F8V}{6250}{1.18}{1.90}{1.18}{-0.07}
    \ROWi{G0V}{6000}{1.05}{1.41}{1.10}{-0.06}
    \ROWi{G2V}{5750}{9.96\EXP{-1}}{1.03}{1.02}{-0.02}
    \ROWi{G5V}{5750}{9.20\EXP{-1}}{8.35\EXP{-1}}{9.20\EXP{-1}}{0.04}
    \ROWi{G8V}{5500}{8.40\EXP{-1}}{6.35\EXP{-1}}{8.77\EXP{-1}}{0.04}
    \ROWi{K0V}{5250}{7.90\EXP{-1}}{4.95\EXP{-1}}{8.50\EXP{-1}}{0.04}
    \ROWi{K1V}{5000}{7.64\EXP{-1}}{3.81\EXP{-1}}{8.22\EXP{-1}}{0.05}
    \ROWi{K2V}{5000}{7.40\EXP{-1}}{3.57\EXP{-1}}{7.95\EXP{-1}}{0.07}
    \ROWi{K3V}{4750}{7.16\EXP{-1}}{2.72\EXP{-1}}{7.69\EXP{-1}}{0.08}
    \ROWi{K4V}{4500}{6.92\EXP{-1}}{2.05\EXP{-1}}{7.44\EXP{-1}}{0.10}
    \ROWi{K5V}{4250}{6.70\EXP{-1}}{1.53\EXP{-1}}{7.20\EXP{-1}}{0.11}
    \ROWi{K7V}{4000}{6.01\EXP{-1}}{1.03\EXP{-1}}{6.69\EXP{-1}}{0.13}
    \ROWi{M0V}{3750}{5.10\EXP{-1}}{6.42\EXP{-2}}{6.00\EXP{-1}}{0.15}
    \ROWi{M1V}{3750}{4.52\EXP{-1}}{5.35\EXP{-2}}{5.48\EXP{-1}}{0.18}
    \ROWi{M2V}{3500}{4.00\EXP{-1}}{3.38\EXP{-2}}{5.00\EXP{-1}}{0.20}
    \ROWi{M3V}{3500}{3.30\EXP{-1}}{2.74\EXP{-2}}{4.50\EXP{-1}}{0.21}
    \ROWi{M4V}{3500}{2.63\EXP{-1}}{1.64\EXP{-2}}{3.49\EXP{-1}}{0.34}
    \hline
  \end{tabular}
\end{table}

\subsection{Stellar Features}
\label{s:stel_feat}

\citet{WoolfPDI-chop} were the first to devise a technique called ``chopping'',
or ``modulation'', enabling nulling interferometers to see only the shot
noise of centrosymetric sources in the FOV (to a first order stellar
leakage, and exozodiacal dust emission).  Given the $10^6$ brightness
dynamic between the star and an exo-Earth, stellar features may introduce
biases in both detection and spectroscopy that cannot be removed by means
of internal modulation.  Such problems are expected from the contribution
of spatially non-symmetric emission features such as non-uniform surface
brightness (\textit{e.g.}  stellar spots, polar caps for fast rotators) or
star misalignment with instrumental line of sight.

\begin{figure}
  \centering
  \includegraphics[width=\columnwidth]{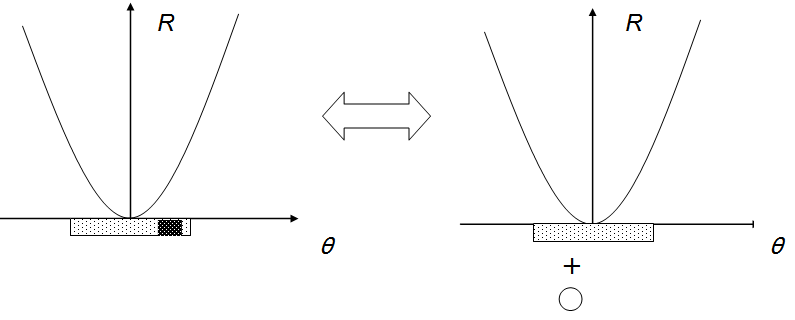}
  \caption{\label{f:spot}Stellar spot - point source equivalence. When viewed through the asymmetric filter that the instrument represents, a stellar spot is equivalent to a symmetrically placed virtual planet. $R$ is the interferometric coronagraphic transmission,
    and $\theta$ is angular sky position parameter.}
\end{figure}

Stellar spots, for instance, represent a localized flux default
(Fig.~\ref{f:spot}).  For a temperature differential of $\sim10^3\,\mathrm{K}$, and a spot area of 10\% of the star's surface, the flux default is $50\,\mathrm{ph}\,\mathrm{s}^{-1}\,\mathrm{m}^{-2}\,\micron^{-1}$.
However, with an ideal nulling transmission on the limb of the star of
$10^{-8}$, this particular feature will not interfere with the signal from
an exo-Earth.

\subsection{Star Output}

\begin{figure}
  \centering
  \includegraphics[width=\columnwidth]{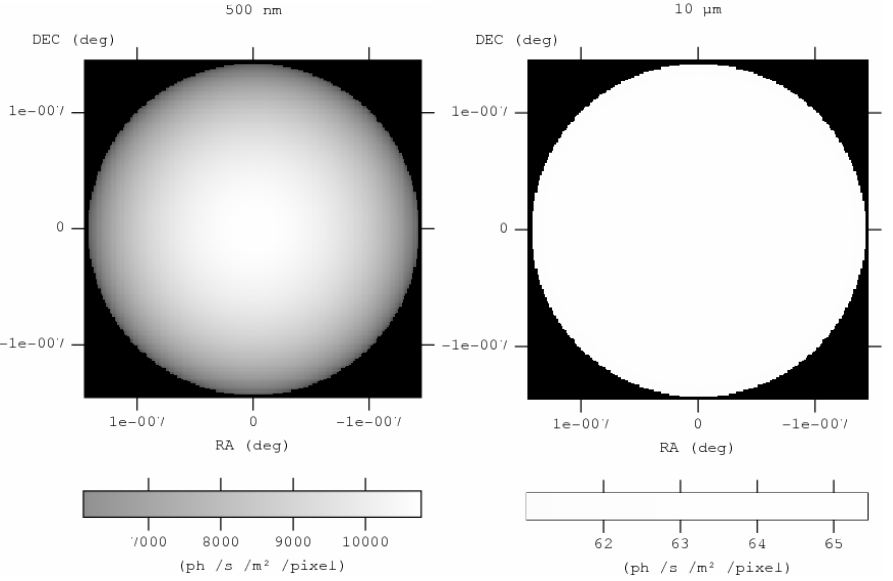}
  \caption{\label{f:origin_star}$209\times209$-pixel \Origin output images
    of a G0V star at 10\,pc at wavelengths $\lambda=0.5\,\micron$ and
    $\lambda=10\,\micron$.  Linear grayscales go from 0 to the maximum
    of brightness in each plot. The bandwidth of the channel is
    1\,\micron.}
\end{figure}

Taking into account the considerations presented in Sects. \ref{s:limb-darkening} through \ref{s:stel_feat}, the option chosen in our
model is to generate a resolved image of the limb-darkened
stellar surface with sufficient resolution to allow for a correct
estimation of the leakage (Fig.~\ref{f:origin_star}).  In order to avoid introducing too many settings in our model (which would prevent inspection of
a wide range of possible scenarios), the only parameter considered  to
account for non-symmetric stellar contribution is the pointing error, i.e. a possible angular offset between the line of sight and the position of
the star.

The images in Fig.~\ref{f:origin_star}, and a majority of the following, are outputs of the widely used \emph{fv} (\texttt{http://heasarc.nasa.gov/ftools/fv}) \Fits viewer and editor program, which is proposed as an interface for the \emph{ORIGIN} software. This was preferred to a dedicated interface development, in order to use tools already familiar to the astrophysical community as much as possible.

\section{Exoplanets}
\label{s:expn}

The exoplanets will be unresolved at the instrument's resolution.  The
specific intensity of the $j$-th planet is therefore:
\begin{equation}
  I_j(\ViewDir,\lambda,t) =
  \Bigl[F_{\mathrm{e},j}(\lambda,t) + F_{\mathrm{r},j}(\lambda,t)\Bigr]
  \,\delta\Bigl(\ViewDir - \ViewDir_j(t)\Bigr)\,,
\end{equation}
where $\delta$ is Dirac's distribution, $\ViewDir_j(t)$ is the angular
position of the planet at date $t$, $F_\RM{e}$ is the specific flux
intrinsically emitted by the planet and $F_\RM{r}$ is the stellar flux
reflected by the planet.  Absorption of the planetary flux by exozodiacal
dust, even in the case of edge-on systems, would assume a global dust
density that would rule out nulling detection, so this is not implemented.
The specific fluxes emitted and reflected by a given planet are discussed
in the following section; we have, however, dropped the planet index $j$ for the sake of simplicity.

\subsection{Planet Emission}
\label{s:franck}

The diversity of exoplanets is expected to be considerable, especially in
the case of terrestrial planets \citep{Gaidos_and_Selsis_2007}.  The
composition of a given terrestrial exoplanet's atmosphere, and thus its
spectrum, will depend on many parameters. Among these parameters are the stellar type and the detailed chemical composition of its parent star, its orbital distance, its
mass, its accretion history, the relative abundance of accreted solids
(silicate, metal, ice) and volatiles (including water), the age of the
system, and possibly, the existence of an extensive biosphere.  Simulation
of planetary atmospheres is a rather young field, and the encompassed
models represent a level of complexity beyond the scope of this work.  The
best approach for an exoplanetary systems FOV model is to be able to use
external spectra, provided by teams working on the computation of realistic
synthetic spectra. A collection of identified object spectra is already
available: phase-dependent Jupiter-like planets at different orbital
distances \citep{Barman_2005}, Earth, Mars or Venus-like planets and derived
terrestrial planets \citep{Selsis_2000, Schindler_2000, Selsis_2002,
2002desmarais, tinetti_earth, TinettiMars}, Earth-replicas orbiting around G,
F, and K stars \citep {Selsis_2000,Segura_2003}, M stars \citep{Segura2005} the
Earth throughout its history \citep{Kaltenegger_2006}, Earth-like planets \New{across} the
habitable zone \citep{these_jimmy}, and "ocean-planets" \citep
{2004ocean_planets}.

Alternatively, it is possible to implement a simple black body emission 
spectral energy distribution. In our model, its effective temperature can be either calculated upon a
radiative equilibrium with the star, or fixed by the user.  The effective
temperature of the planet $T_\RM{p}$ derived from radiative equilibrium
with the star is:
\begin{equation}
  T_\RM{p} = (1 - A_\RM{p})^{1/4}
  \, \sqrt{\frac{\Rstar}{2\,r}} \,\Tstar\,,
  \label{eq:planet-equilibrium-temperature}
\end{equation}
where $A_\RM{p}$ is the Bond albedo, and the distance $r$ from the planet to
the star depends on the observation date $t$ and is computed by solving the
orbital equations.  If a simple black body emission is used, the emitted
flux is
\begin{equation}
  F_\RM{e}(\lambda,t) =
  \pi\,\left(\frac{R_\RM{p}}{d}\right)^2\,B_{\lambda}(T_\RM{p})\,,
\end{equation}
where $B_{\lambda}$ is the Planck function.

\subsection{Reflected Flux}
\label{s:rflctd}

Depending on their albedo and on their phase with respect to the observer,
the planets partially reflect the flux of their hosting star.  The specific
flux received by a planet from its star, at a distance $r$ from the star
is:
\begin{equation}
  F_\lambda(r)
  = 2\,\pi\,\left(\frac{R_\star}{r}\right)^2\,\Inorm\,Q_\lambda\,,
  \label{e:star-flux}
\end{equation}
where the factor $Q_\lambda$ depends on the star limb-darkening law:
\begin{equation}
  Q_\lambda = \int_0^1 D_\lambda(\mu)\,\mu\,\mathrm{d}\mu\,.
\end{equation}
The factor $Q_\lambda$ can be computed for the considered limb-darkening laws:
\begin{equation}
  Q_\lambda = \left\{
  \begin{array}{ll}
    1/2 & \mbox{\small(black body)}\\
    (3 - x_\lambda)/6 & \mbox{\small(linear law)}\\
    (9 - 3\,x_\lambda + 2\,y_\lambda)/18
    & \mbox{\small(logarithmic law)}\\
    (15 - 5\,x_\lambda - 3\,y_\lambda)/30
    & \mbox{\small(square-root law)}\\
  \end{array}\right.
\end{equation}
The flux reflected by the planet then is:
\begin{equation}
  F_\RM{r}(\lambda,t) =
  \mathcal{R}(\lambda,\xi)\,\left(\frac{R_\RM{p}}{d}\right)^2
  \,F_{\lambda}(r)\,.
\end{equation}
$F_{\lambda}(r)$ is the specific flux received by the planet from its hosting star as given by
Eq.~(\ref{e:star-flux}) and $d$ is its
distance from the observer. $\mathcal{R}(\lambda, \xi)$ is the phase-dependent reflectivity of the planet, as seen at an angle $\xi$ from the star. This general expression of the reflectivity enables our model (which is open) to be updated with expressions taking into account specular reflection on a planet's surface, as well as Mie and Rayleigh diffusion, which is beyond the scope of our present work. For now, we have used a Lambertian model $Ref(\lambda,\xi) = A_{\lambda}\,\phi(\xi)$. $A_{\lambda}$ is a constant and $\phi(\xi)$ is the fraction of the planetary disk
surface illuminated by the star as seen by the observer (phase):

\begin{equation}
  \phi(\xi) = \frac{1}{2}\,
      \left[1 + \sin(\pi-\xi)\right]
      = \frac{1}{2}\,
      \left[1 + \sin(\nu(t) + \omega(t))\,\sin i\right]
\end{equation}
where $i$ is the inclination of the orbit, $\nu$ is the true anomaly, and
$\omega$ is the argument of the periastron of the considered planet at the
time of observation.

\begin{figure}
  \includegraphics[width=\columnwidth]{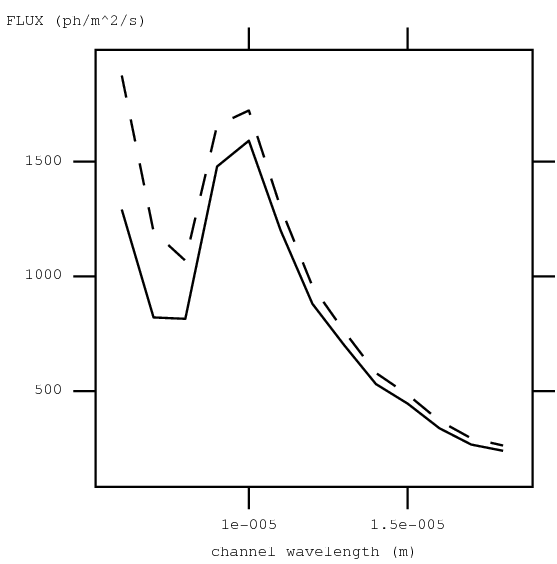}
  \caption{\label{f:reflection}Full night side (solid) and full day side
    (dashed) emissions. Night: emission model of a 1.1\,R$_{\RM{Jupiter}}$
    planet of albedo 0.1, orbiting at 0.02\,AU from parent star from
    \citet{Barman_2005}. Day: added full-disk reflection of Kurucz model
     120 (G0V) stellar flux, from \citet{VanHamme}. All seen at
    10\,pc. Channel width: 1~\micron.}
\end{figure}

Figure \ref{f:reflection} exemplifies the importance of this phenomenon as
simulated with our model. Actually, the emission spectrum of
planets so close to their star is not uniform with their apparent phase
\citep{HarringtonHotspot}.

\section{Other Sources}

\subsection{Local Zodiacal Cloud}
\label{s:local_zodi}

\begin{figure}
  \resizebox{\hsize}{!}{\includegraphics{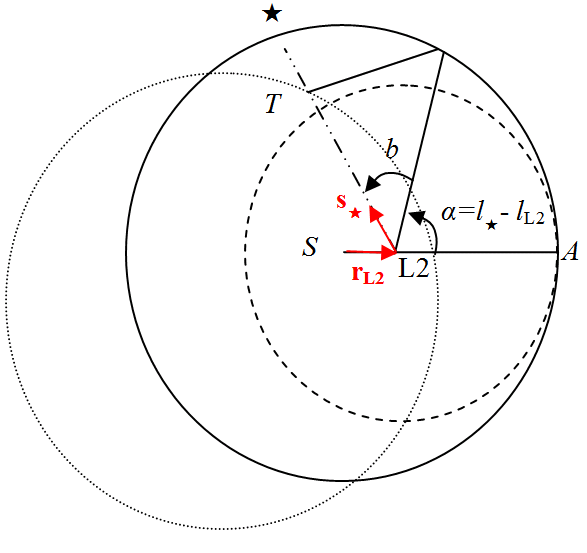}}
  \caption{\label{f:local_zodi}How the optical depth, for a target at a
      given ecliptical latitude $b$, increases with the target's
      L2-relative ecliptical longitude $\alpha$ (S-Sun, L-L2 point(Sun -
      Earth), A-Antisolar direction, $l$-ecliptical longitude).}
\end{figure}

In order to compute the local zodiacal dust emission, we integrate along
the line of sight (Fig.~\ref{f:local_zodi}), from the observer's location
towards the target, through a 3-dimensional sampling of the Kelsall
model \citep{Kelsall}. Our approach is thus similar to that of \citet{landgraf_jehn}, with some minor differences, and from a "background drift",  fixed pointing direction perspective. Since there is no theoretical outer limit to the exponential law
of the dust density in this model we integrate to the outer limit of the
physical cloud; i.e. its collisional origin in the asteroid belt, at SA
$\approx 5\,\RM{AU}$.

Given the size of the interferometer's FOV, and with respect to the value
of this outer limit, the local zodiacal flux  contributes uniformly to
the image. This is equivalent to a global noise level in the nulling data,
depending only on the target's sky position and on the instrument's position
at a given observation date.

Since scattering by dust is negligible compared to the thermal emission at
10~\micron, it can easily be seen that it is best, at a given mission time, to
observe targets in the antisolar ecliptic meridian, because that is where the
optical depth of the cloud is minimal. However, during spectroscopic
observations, which can be 4 months long, the optical depth along the line
of sight will vary.

Let $\ViewDir_\star$ be the direction of the target star (the time
parameter $t$ is omitted for the sake of readability):
\begin{equation}
  I_\LocalZodi(\ViewDir_\star,\lambda) =
  \int_{0}^{\RM{L2\,T}}
  J_\LocalZodi(\BF{r}_\RM{L2} + u\,\ViewDir_\star,\lambda) \,\RM{d}u
\end{equation}
where the position $\BF{r}=\BF{r}_\RM{L2} + u\,\ViewDir_\star$ is relative
to the Sun, $\BF{r}_\RM{L2}$ is the observer position, $u$ is the
distance from the observer towards the source along the line of sight, and
where $J_\LocalZodi$ is the emitted specific intensity per unit of depth,
i.e. the power emitted by the dust at position $\BF{r}$, per unit of
volume, per unit of solid angle, per unit of spectral bandwidth, given by
the model of \citet{Kelsall}.

The direction of observation $\ViewDir_\star$ is $(\RA,\DEC)$ in equatorial
coordinates and, to perform the integration, it must be converted into
ecliptic coordinates $(l,b)$.

\begin{figure}
  \resizebox{\hsize}{!}{\includegraphics{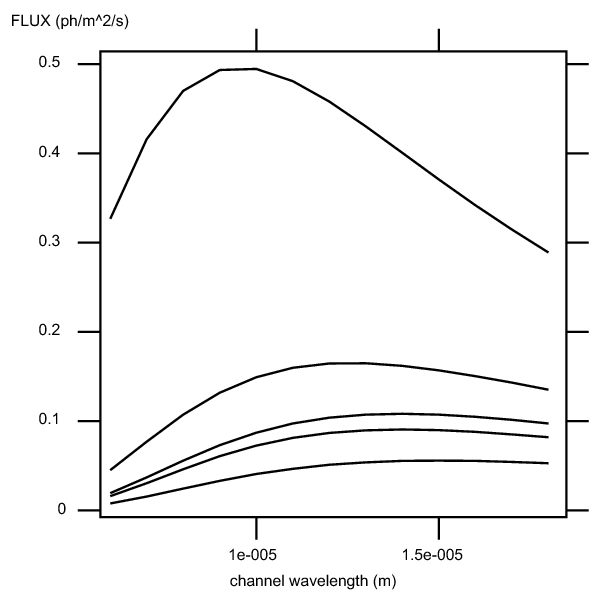}}
  \caption{\label{f:local_result}Local zodiacal cloud emission drift at L2
    point over an evenly spread interval of 6 months, steadily increasing
    from the anti-solar direction (lowest) to solar (highest). The
    ecliptical latitude of the target is 10$^{\circ}$, channel width is
    1~\micron. FOV radius is $0\farcs6$.}
\end{figure}

Figure \ref{f:local_result} shows the drift of the local zodiacal emission background, for a given target, from the L2 point, over
a evenly spread period of 6 months.  At around 14~\micron the zodiacal
drift over 3~months\footnote{Collectors can point no further than $\sim80\degr$ from the antisolar direction to enable solar shielding} is
200\%. Thus the global noise level drift is $\approx14\,\%$ in that
particular channel, over a duration not unlikely for a spectroscopic
observation.

\subsection{The Exozodiacal Cloud}
\label{s:exo-zodi}

In our model, we assume that the exozodiacal cloud dust has similar
properties to the solar zodiacal cloud (Sect.~ \ref{s:local_zodi}), hence we
simulate it by scaling the model of \citet{Kelsall}. All
parameters are free, enabling for instance to produce large clumps of dust
useful for robustness testing of planet signal extraction algorithms.

\subsection{Galactic Interstellar Medium}
\label{s:ISM}

\paragraph{Emission levels.}

Galactic interstellar medium (ISM) IR emission can reach hundreds of
$\mathrm{MJy}\,\mathrm{sr}^{-1}$ in the infrared \citep{Schlegel}.  Over the typical $0\farcs6$ field of view of
interferometer, this equates to 2500\,$\mathrm{ph}\,\mathrm{s}^{-1}\,\mathrm{m}^{-2}\,\micron^{-1}$ at 10~\micron. We carried
out simulations on ESA code \citep{darwinsim}, showing that the additional noise
level provided by a such a background doubles the detection time of an
exo-Earth.

\begin{figure}
  \centering
  \includegraphics[width=\columnwidth]{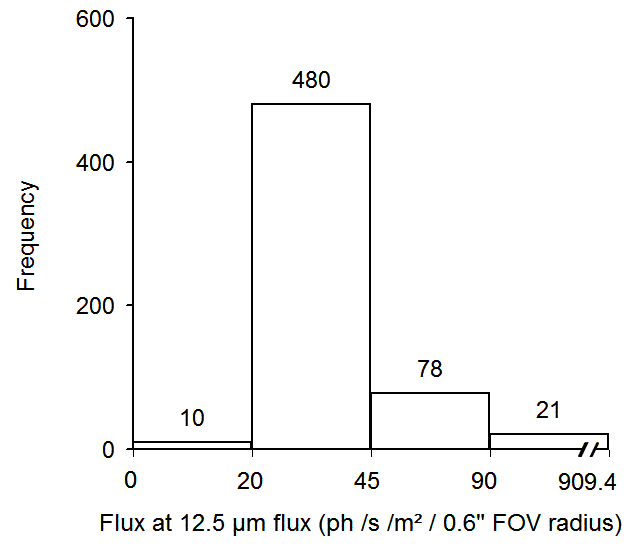}
  \caption{\label{f:hist}Histogram of galactic interstellar medium
    emission at 12.5~\micron, for 609 targets form
    \cite{Kaltenegger}}
\end{figure}

We have therefore extracted from the \textit{Improved Reprocessing of the IRAS
Survey (IRIS)} maps \citep{Miville} 12.5 and 25~\micron background
emission levels for the target list of
\citet{Kaltenegger}. Figure~\ref{f:hist} shows the histogram for the
emission at 12.5~\micron.  Background emissions (again, over the FOV) do not
exceed 900~$\mathrm{ph}\,\mathrm{s}^{-1}\,\mathrm{m}^{-2}\,\micron^{-1}$,
with a mean value of $\sim41$ and a median value of
$\sim32$.

It is worth noting that the resolution of the \textit{IRIS} maps is $3\farcm8 $. How do we know
that in such a pixel, we do not include the flux of unresolved sources
(galactic stars, background galaxies, etc.), thus overestimating the flux of
the ISM that would actually be intercepted by the interferometer's FOV? Typical star count
values for magnitudes greater that 20 (K band) range from $3[\log N
\mathrm{deg}^{-2}\,0.5 \mathrm{mag}^{-1}]$ for the galactic disk (not in the galactic
plane) according to \citet{Girardi} to 4.5 for the bulge
\citep{Rodgers}; this is equivalent to 25-50 stars per 0.5 mag per \textit{IRIS} pixel. We can therefore
keep in mind that the values of the ISM we have extracted may be
conservative. However, given the above-mentioned statistics on current
candidate lists, we have not proceeded to implement this correction yet.

\paragraph{Spectral variability.}

As can be derived from \citet[Fig.~1 therein]{Verstraete}, variations of
the continuum emission of the ISM, around 18~\micron and integrated over
the $0\farcs6$ FOV of the instrument, can reach a photon noise contribution
of 50\,$\mathrm{ph}\,\mathrm{s}^{-1}\,\mathrm{m}^{-2}\,\micron^{-1}$. ISM background calibration
is thus required for spectroscopy.

The reddening absorption due to interstellar molecular clouds between ourselves
and the targets, given their relative proximity, is too faint to be a bias.

\paragraph{Variability over the FOV.}

Diffuse background ISM structure studies have a resolution limit of 10\arcsec \citep{Ingalls}. As can be seen in Fig.~3 of this reference,
extrapolation of the $\alpha = -3.5$ power law to the FOV spatial frequency
($0\farcs6\leftrightarrow1.6\,\RM{arcsec}^{-1}$) leads to an
extrapolated power level of $10^{-7}\,\RM{MJy}^2\,\RM{sr}^{-1}$,
corresponding to a statistical flux variation amplitude over a $0\farcs6$
FOV of $10^{-9}~$Jy at 24~\micron, or
$6\times10^{-4}~\mathrm{ph}\,\mathrm{s}^{-1}\,\mathrm{m}^{-2}\,\micron^{-1}$.
This is comparable to the flux of an exo-Earth (typically
$0.5~\mathrm{ph}\,\mathrm{s}^{-1}\,\mathrm{m}^{-2}\,\micron^{-1}$
at 10~\micron). The emission of the cold ISM at 10~\micron should be even
lower, so the variation of the ISM emission in the FOV should not be
visible in the nulling data processing, at exo-Earth detection
level. Finally, it can be noted that the parallax of the closest targets
($0\farcs2$) is only a fraction of the FOV, so there should be no galactic
ISM background drift with the parallax.

\subsection{Faint Background Objects}
\label{s:bkgrd_obj}

\begin{figure}
  \centering
  \includegraphics[width=\columnwidth]{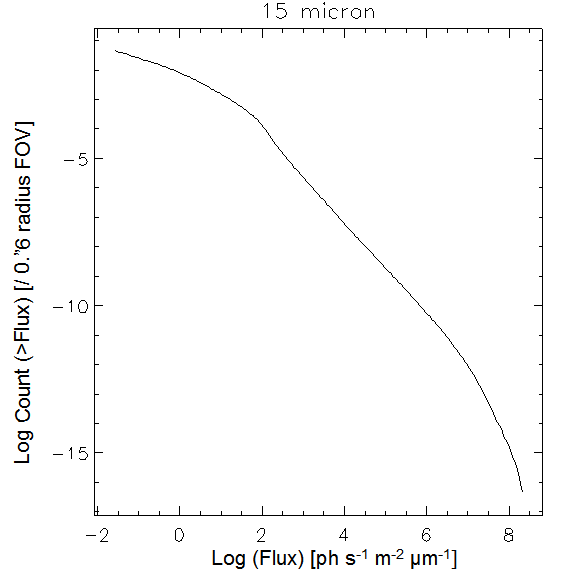}
  \caption{\label{f:gal_count_int}Integrated field galaxy counts. Plot has
    similar profile at 60 and 170\,\micron, with maximums steadily decreasing  with the wavelength.}
\end{figure}

Figure \ref{f:gal_count_int} displays an integration of a previously
established galaxy count histogram \citep{Lagache}.  As mentioned above, an
exo-Earth at 10~pc emits
$0.5~\mathrm{ph}\,\mathrm{s}^{-1}\,\mathrm{m}^{-2}\,\micron^{-1}$ at
10~\micron.  We see that, statistically, there should be 0.01 objects
brighter than that in a $0\farcs6$ FOV.
In order to examine galactic stars in the FOV, we can investigate further the
reference cited in Sect.~\ref{s:ISM} \citep[Fig.~6 therein]{Girardi}. Let us consider the star count peak in 
the K band (2 \micron, longest wavelength available in this reference; in the FOV, that represents 2.2~$\times~10^{-5}$ objects per 0.5 mag bandwidth, so it is not even necessary to consider the integrated count.  We conclude that it is
unlikely that one will encounter background stars emitting at the level of an exo-Earth or greater in the FOV. Were faint background objects still to occur in the FOV, they could be easily ruled out spectroscopically, and by monitoring their orbital motion during revisits.

In addition to the final remark of Sect.~\ref{s:ISM} concerning the ISM emission, the
numbers above remain negligible if an extended FOV (taking into account the
parallax) is considered instead, so we have not accounted for time
variability of $I_\BG$ (galactic background drift).

Because of the above, our model does not specifically implement FOV background stars or galaxies.

\section{Discussion}

It is interesting to consider whether one or several of the source features
identified above was already an ``astrophysical limitation'' in achieving the scientific objective of these exo-Earth finding missions. We recall that FOV source features where screened on the intuitive criterion whereby ``anything producing a signal at the level of an exo-Earth deserves modeling'', in the prospect of full
end-to-end testing simulations.

Currently, target catalogs for these survey missions have been started; for now, these are based on astrobiological
interest and the few informed astro-engineering requirements available
today, such as the existence of a secondary star too near and bright \citep{Kaltenegger}\footnote{\raggedright Also, recent unpublished work by Brown et al. \texttt{http://sco.stsci.edu/tpf\_tldb/downloads/}
\texttt{TPF\_SWG\_presentation\_feb24\_04.pdf}}. Surveys of exozodiacal
dust levels of nearby stars are planned or in
progress\footnote{\raggedright Traub and Kuchner, Shared Risk Keck Nuller
observations \texttt{http://planetquest.jpl.nasa.gov/Navigator/}
\texttt{keck\_sharedrisk.cfm\#traub}}; the statistics of this unknown
parameter alone may impose dramatic revision of the scale and cost of missions, even though elements in favor of optimism do 	exist
\citep{BeichmanDisks2006}. 

In this context, the features of the sources described here are, in most cases, an order of magnitude below the exozodiacal
cloud issue. Beyond the classical limitation of shot noise from
unsought-for sources, additional requirements on the integration time
(hence on the cost) may arise from signal processing specificities, such \New{as}
systematic multiple planet signal disentangling, regardless of their
relative positions and kinematics. This is the purpose of our ongoing work.

\section{Summary}
\label{s:summary}

We have defined a FOV physical model of exoplanetary system scenes, and
proposed an associated \Fits input and output format.  The input format is
a tentative standard for defining exoplanetary systems. The output format
is an input format for exoplanet seeking instrumental simulators. They are
both described in the Appendices.

Submitting a fixed resolution image to an instrumental simulator is not
practical, given the dynamics in resolution and flux between the
various typical sources. The output flux is modeled by a ``layering'' of various sources
(star, planets, dust). Each ``layer'' is of one type among i) a 3-D spectral flux image (for resolved sources), ii) a spectral flux and a position (for unresolved
sources), iii) a FOV-uniform spectral contribution. For each source we have
examined the detectable features that need to be modeled.

Depending on the spectral type of the parent star and of the wavelength,
omitting to model limb-darkening is shown to induce a bias in the
estimation of the leakage noise of up to $\sim35\,\%$.  Stellar spots produce a fainter signal than that of an exo-Earth.  Local zodiacal drift is found to be
smaller than 14\,\% in any given spectral channel.  Galactic ISM background for a
current list of $\sim600$ targets peaks at
$900\,\RM{ph}\,\RM{s}^{-1}\,\RM{m}^{-2}\,\micron^{-1}$, with a mean value
of $\sim41$ and a median value of $\sim32$ for a typical FOV of $0\farcs6$.
Simulations show that a doubling of the detection time of an exo-Earth is
induced by a $2500\,\RM{ph}\,\RM{s}^{-1}\,\RM{m}^{-2}\,\micron^{-1}$
galactic emission background (compared to no background emission, this is the order of magnitude of the highest emissions). Finally, background galactic stars and distant galaxies as bright as, or brighter than an exo-Earth are unlikely in the FOV, even extended by parallax drift.

The model specifications have been embedded by Starlab \& Thales Alenia
Space (formerly Alcatel Alenia Space, see acknowledgments) into a Java simulator called \Origin, soon to be open-source, using the input/output definition standards detailed in the
Appendices.  We are in the process of building upon this work to obtain an end-to-end simulation approach.

\begin{acknowledgements}

Part of this work was conducted in the frame of the \emph{Reconstruction of
ExoSolar System Properties} (RESSP) study for ESA/ESTEC contract
18701/04/NL/HB, led by Thales Alenia Space. A. Belu was supported at the time of this work by a
\textit{Centre National de la Recherche Scientifique} grant. The authors
acknowledge useful discussions with members of the MATIS Team, LUAN, and thank the referee, Wesley Traub, for the attentive review of the manuscript.
\end{acknowledgements}

\bibliographystyle{aa} 
\bibliography{origin} 

\begin{thebibliography}{46}
\expandafter\ifx\csname natexlab\endcsname\relax\def\natexlab#1{#1}\fi

\bibitem[{Absil(2001)}]{Absil-2001-thesis-nulling_interferometry}
Absil, O. 2001, PhD thesis, Universite de Liége

\bibitem[{{Absil} {et~al.}(2006){Absil}, {den Hartog}, {Gondoin}, {Fabry},
  {Wilhelm}, {Gitton}, \& {Puech}}]{Absil_ground_AA}
{Absil}, O., {den Hartog}, R., {Gondoin}, P., {et~al.} 2006, \aap, 448, 787

\bibitem[{{Barman} {et~al.}(2005){Barman}, {Hauschildt}, \&
  {Allard}}]{Barman_2005}
{Barman}, T.~S., {Hauschildt}, P.~H., \& {Allard}, F. 2005, \apj, 632, 1132

\bibitem[{{Beichman} {et~al.}(2006){Beichman}, {Bryden}, {Stapelfeldt},
  {Gautier}, {Grogan}, {Shao}, {Velusamy}, {Lawler}, {Blaylock}, {Rieke},
  {Lunine}, {Fischer}, {Marcy}, {Greaves}, {Wyatt}, {Holland}, \&
  {Dent}}]{BeichmanDisks2006}
{Beichman}, C.~A., {Bryden}, G., {Stapelfeldt}, K.~R., {et~al.} 2006, \apj,
  652, 1674

\bibitem[{{Beichman} {et~al.}(1999){Beichman}, {Woolf}, \&
  {Lindensmith}}]{BeichmanJPLBook99}
{Beichman}, C.~A., {Woolf}, N.~J., \& {Lindensmith}, C.~A., eds. 1999, {The
  Terrestrial Planet Finder (TPF) : a NASA Origins Program to search for
  habitable planets} (The TPF Science Working Group National Aeronautics and
  Space Administration ; Pasadena, Calif.~: Jet Propulsion Laboratory,
  California Institute of Technology, (JPL publication ; 99-3))

\bibitem[{{Bracewell}(1978)}]{Bracewell}
{Bracewell}, R.~N. 1978, \nat, 274, 780

\bibitem[{den Hartog(2006)}]{darwinsim}
den Hartog, R. 2006, private communication

\bibitem[{{Des Marais} {et~al.}(2002){Des Marais}, {Harwit}, {Jucks},
  {Kasting}, {Lin}, {Lunine}, {Schneider}, {Seager}, {Traub}, \&
  {Woolf}}]{2002desmarais}
{Des Marais}, D.~J., {Harwit}, M.~O., {Jucks}, K.~W., {et~al.} 2002,
  Astrobiology, 2, 153

\bibitem[{{Draper} {et~al.}(2006){Draper}, {Elias}, {Noecker}, {Dumont}, {Lay},
  \& {Ware}}]{Draper}
{Draper}, D.~W., {Elias}, II, N.~M., {Noecker}, M.~C., {et~al.} 2006, \aj, 131,
  1822

\bibitem[{{Ferrari} {et~al.}(2006){Ferrari}, {Carbillet}, {Serradel}, {Aime},
  \& {Soummer}}]{FerrariUAI}
{Ferrari}, A., {Carbillet}, M., {Serradel}, E., {Aime}, C., \& {Soummer}, R.
  2006, in IAU Colloq. 200: Direct Imaging of Exoplanets: Science {\&}
  Techniques, ed. C.~{Aime} \& F.~{Vakili}, 565--570

\bibitem[{Fliegel \& van Flandern(1968)}]{Fliegel:vanFlandern:1968}
Fliegel, H.~F. \& van Flandern, T.~C. 1968, Communications of the ACM, 11

\bibitem[{{Fridlund}(2000)}]{FridlundESA2000}
{Fridlund}, M. 2000, in ESA SP-451: Darwin and Astronomy : the Infrared Space
  Interferometer, ed. B.~{Sch{\"u}rmann}, 11--+

\bibitem[{{Gaidos} \& {Selsis}(2007)}]{Gaidos_and_Selsis_2007}
{Gaidos}, E. \& {Selsis}, F. 2007, in Protostars and Planets V, B. Reipurth, D.
  Jewitt, and K. Keil (eds.), University of Arizona Press, Tucson, 951 pp.,
  2007., p. 929-944, ed. B.~{Reipurth}, D.~{Jewitt}, \& K.~{Keil}, 929--944

\bibitem[{{Girardi} {et~al.}(2005){Girardi}, {Groenewegen}, {Hatziminaoglou},
  \& {da Costa}}]{Girardi}
{Girardi}, L., {Groenewegen}, M.~A.~T., {Hatziminaoglou}, E., \& {da Costa}, L.
  2005, A\&A, 436, 895

\bibitem[{Hanisch {et~al.}(2001)Hanisch, Farris, Greisen, Pence, Schlesinger,
  Teuben, Thompson, \& Warnock}]{FITS}
Hanisch, R.~J., Farris, A., Greisen, E.~W., {et~al.} 2001, A\&A, 376, 359

\bibitem[{{Harrington} {et~al.}(2006){Harrington}, {Hansen}, {Luszcz},
  {Seager}, {Deming}, {Menou}, {Cho}, \& {Richardson}}]{HarringtonHotspot}
{Harrington}, J., {Hansen}, B.~M., {Luszcz}, S.~H., {et~al.} 2006, Science,
  314, 623

\bibitem[{{Ingalls} {et~al.}(2004){Ingalls}, {Miville-Desch{\^e}nes}, {Reach},
  {Noriega-Crespo}, {Carey}, {Boulanger}, {Stolovy}, {Padgett}, {Burgdorf},
  {Fajardo-Acosta}, {Glaccum}, {Helou}, {Hoard}, {Karr}, {O'Linger}, {Rebull},
  {Rho}, {Stauffer}, \& {Wachter}}]{Ingalls}
{Ingalls}, J.~G., {Miville-Desch{\^e}nes}, M.-A., {Reach}, W.~T., {et~al.}
  2004, ApJ S, 154, 281

\bibitem[{{Kaltenegger} {et~al.}(2006){Kaltenegger}, {Eiroa}, {Stankov}, \&
  {Fridlund}}]{Kaltenegger}
{Kaltenegger}, L., {Eiroa}, C., {Stankov}, A., \& {Fridlund}, M. 2006, in
  Direct Imaging of Exoplanets: Science {\&} Techniques. Proceedings of the IAU
  Colloquium \#200, Edited by C. Aime and F. Vakili. Cambridge, UK: Cambridge
  University Press, 2006., pp.89-92, ed. C.~{Aime} \& F.~{Vakili}, 89--92

\bibitem[{{Kaltenegger} {et~al.}(2007){Kaltenegger}, {Traub}, \&
  {Jucks}}]{Kaltenegger_2006}
{Kaltenegger}, L., {Traub}, W.~A., \& {Jucks}, K.~W. 2007, \apj, 658, 598

\bibitem[{{Kelsall} {et~al.}(1998){Kelsall}, {Weiland}, {Franz}, {Reach},
  {Arendt}, {Dwek}, {Freudenreich}, {Hauser}, {Moseley}, {Odegard},
  {Silverberg}, \& {Wright}}]{Kelsall}
{Kelsall}, T., {Weiland}, J.~L., {Franz}, B.~A., {et~al.} 1998, ApJ, 508, 44

\bibitem[{{Lagache} {et~al.}(2004){Lagache}, {Dole}, {Puget},
  {P{\'e}rez-Gonz{\'a}lez}, {Le Floc'h}, {Rieke}, {Papovich}, {Egami},
  {Alonso-Herrero}, {Engelbracht}, {Gordon}, {Misselt}, \&
  {Morrison}}]{Lagache}
{Lagache}, G., {Dole}, H., {Puget}, J.-L., {et~al.} 2004, ApJ S, 154, 112

\bibitem[{Landgraf(2004)}]{landgraf_orbit}
Landgraf, M. 2004, DARWIN Mission Analysis: Operational Phase and Transfer, MAO
  Working Paper 480, European Space Agency, Directorate of Technical and
  Operational Support, Ground System Engineering Department, Mission Analysis
  Office

\bibitem[{{Landgraf} \& {Jehn}(2001)}]{landgraf_jehn}
{Landgraf}, M. \& {Jehn}, R. 2001, \apss, 278, 357

\bibitem[{Lay(2004)}]{lay_2004}
Lay, O.~P. 2004, Applied Optics, 43, 6100

\bibitem[{{Leger} {et~al.}(1996){Leger}, {Mariotti}, {Mennesson}, {Ollivier},
  {Puget}, {Rouan}, \& {Schneider}}]{LegerIcar96}
{Leger}, A., {Mariotti}, J.~M., {Mennesson}, B., {et~al.} 1996, Icarus, 123,
  249

\bibitem[{{L\'eger} {et~al.}(2004){L\'eger}, {Selsis}, {Sotin}, {Guillot},
  {Despois}, {Mawet}, {Ollivier}, {Lab\`eque}, {Valette}, {Brachet},
  {Chazelas}, \& {Lammer}}]{2004ocean_planets}
{L\'eger}, A., {Selsis}, F., {Sotin}, C., {et~al.} 2004, Icarus, 169, 499

\bibitem[{{Marsh} {et~al.}(2006){Marsh}, {Velusamy}, \& {Ware}}]{Marsh}
{Marsh}, K.~A., {Velusamy}, T., \& {Ware}, B. 2006, \aj, 132, 1789

\bibitem[{{Miville-Desch{\^e}nes} \& {Lagache}(2005)}]{Miville}
{Miville-Desch{\^e}nes}, M.-A. \& {Lagache}, G. 2005, ApJ S, 157, 302

\bibitem[{{Mugnier} {et~al.}(2006){Mugnier}, {Thi{\'e}baut}, \&
  {Belu}}]{MugnierITHD}
{Mugnier}, L., {Thi{\'e}baut}, E., \& {Belu}, A. 2006, in EAS Publications
  Series, 69--83

\bibitem[{{Paillet}(2006)}]{these_jimmy}
{Paillet}, J. 2006, PhD thesis, Universit\'{e} Paris XI and Ecole Normale
  Sup\'{e} rieure de Lyon

\bibitem[{{Rodgers} {et~al.}(1986){Rodgers}, {Harding}, \& {Ryan}}]{Rodgers}
{Rodgers}, A.~W., {Harding}, P., \& {Ryan}, S. 1986, AJ, 92, 600

\bibitem[{{Schindler} \& {Kasting}(2000)}]{Schindler_2000}
{Schindler}, T.~L. \& {Kasting}, J.~F. 2000, Icarus, 145, 262

\bibitem[{{Schlegel} {et~al.}(1998){Schlegel}, {Finkbeiner}, \&
  {Davis}}]{Schlegel}
{Schlegel}, D.~J., {Finkbeiner}, D.~P., \& {Davis}, M. 1998, ApJ, 500, 525

\bibitem[{{Schmidt-Kaler}(1982)}]{SK}
{Schmidt-Kaler}, T. 1982, Bulletin d'Information du Centre de Donnees
  Stellaires, 23, 2

\bibitem[{{Segura} {et~al.}(2005){Segura}, {Kasting}, {Meadows}, {Cohen},
  {Scalo}, {Crisp}, {Butler}, \& {Tinetti}}]{Segura2005}
{Segura}, A., {Kasting}, J.~F., {Meadows}, V., {et~al.} 2005, Astrobiology, 5,
  706

\bibitem[{{Segura} {et~al.}(2003){Segura}, {Krelove}, {Kasting}, {Sommerlatt},
  {Meadows}, {Crisp}, {Cohen}, \& {Mlawer}}]{Segura_2003}
{Segura}, A., {Krelove}, K., {Kasting}, J.~F., {et~al.} 2003, Astrobiology, 3,
  689

\bibitem[{{Selsis}(2000)}]{Selsis_2000}
{Selsis}, F. 2000, in ESA SP-451: Darwin and Astronomy : the Infrared Space
  Interferometer, ed. B.~{Sch{\"u}rmann}, 133--140

\bibitem[{{Selsis} {et~al.}(2002){Selsis}, {Despois}, \&
  {Parisot}}]{Selsis_2002}
{Selsis}, F., {Despois}, D., \& {Parisot}, J.-P. 2002, AA, 388, 985

\bibitem[{{Thiébaut} {et~al.}(2007){Thiébaut}, {Mugnier}, \&
  {Belu}}]{Thiebaut2007}
{Thiébaut}, E., {Mugnier}, L., \& {Belu}, A. 2007, in preparation

\bibitem[{{Thi{\'e}baut} \& {Mugnier}(2006)}]{Thiebaut}
{Thi{\'e}baut}, E. \& {Mugnier}, L. 2006, in Direct Imaging of Exoplanets:
  Science Techniques. Proceedings of the IAU Colloquium \#200, Edited by C.
  Aime and F. Vakili. Cambridge, UK: Cambridge University Press, 2006.,
  pp.547-552, ed. C.~{Aime} \& F.~{Vakili}, 547--552

\bibitem[{{Tinetti} {et~al.}(2005){Tinetti}, {Meadows}, {Crisp}, {Fong },
  {Velusamy}, \& {Snively}}]{TinettiMars}
{Tinetti}, G., {Meadows}, V.~S., {Crisp}, D., {et~al.} 2005, Astrobiology, 5,
  461

\bibitem[{{Tinetti} {et~al.}(2006){Tinetti}, {Meadows}, {Crisp}, {Kiang},
  {Kahn}, {Fishbein}, {Velusamy}, \& {Turnbull}}]{tinetti_earth}
{Tinetti}, G., {Meadows}, V.~S., {Crisp}, D., {et~al.} 2006, Astrobiology, 6,
  881

\bibitem[{{Traub} {et~al.}(2006){Traub}, {Levine}, {Shaklan}, {Kasting},
  {Angel}, {Brown}, {Brown}, {Burrows}, {Clampin}, {Dressler}, {Ferguson},
  {Hammel}, {Heap}, {Horner}, {Illingworth}, {Kasdin}, {Kuchner}, {Lin},
  {Marley}, {Meadows}, {Noecker}, {Oppenheimer}, {Seager}, {Shao},
  {Stapelfeldt}, \& {Trauger}}]{TraubTPFC_SPIE2006}
{Traub}, W.~A., {Levine}, M., {Shaklan}, S., {et~al.} 2006, in Advances in
  Stellar Interferometry. Edited by Monnier, John D.; Sch{\"o}ller, Markus;
  Danchi, William C.. Proceedings of the SPIE, Volume 6268, pp. (2006).

\bibitem[{Van~Hamme(1993)}]{VanHamme}
Van~Hamme, W. 1993, AJ, 106

\bibitem[{{Verstraete} {et~al.}(2001){Verstraete}, {Pech}, {Moutou},
  {Sellgren}, {Wright}, {Giard}, {L{\'e}ger}, {Timmermann}, \&
  {Drapatz}}]{Verstraete}
{Verstraete}, L., {Pech}, C., {Moutou}, C., {et~al.} 2001, A\&A, 372, 981

\bibitem[{{Woolf} {et~al.}(1998){Woolf}, {Angel}, {Beichman}, {Burge}, {Shao},
  \& {Tenerelli}}]{WoolfPDI-chop}
{Woolf}, N.~J., {Angel}, J.~R.~P., {Beichman}, C.~A., {et~al.} 1998, in Proc.
  SPIE Vol. 3350, Astronomical Interferometry, ed. R.~D. {Reasenberg}, 683--689

\end{thebibliography}

\begin{appendix}

\section{Input format}
\label{s:input-format}

We have developed a \Fits \citep{FITS} standard specifying the input
parameters for modeling an exoplanetary system.  This
format takes advantage of the building block structure of \Fits files.
This building block approach also enables modular storage of stereotypes,
as well as the possibility of linking to
exterior databases in the future.

\subsection{Overview and Primary HDU}

All the data necessary to define a FOV (that can be read by flux
calculators such as the \Origin software -- Sect.~\ref{s:summary}) are stored
in a \Fits file. Databases, such as chromatic specific intensities of
bodies, are stored in \Fits binary tables (\textit{i.e.}, with
\texttt{XTENSION='BINTABLE'}), whereas scalar parameters are stored in the
headers of the extensions.  The primary	 Header Data Unit (HDU) of an
\Origin input data file is informational only and contains no
data\footnote{According to \Fits standard, primary header can only contain
image data, not binary tables.}.  An example of such a primary HDU is
provided in Table~\ref{tab:primary-hdu}.

An \Origin input data file provides the following HDUs:
\begin{itemize}
\item general scenario parameters (spectral channels, observation epoch,
  duration of the observation and number of snapshots to be produced during
  that time frame, FOV resolution and sizes),
\item star parameters,
\item local and exozodiacal cloud parameters,
\item planet(s) parameters.
\end{itemize}
The order of HDUs is irrelevant, the file contains at most one HDU of each
type (except for planets).  HDUs are identified by their names (value of
\texttt{EXTNAME} \Fits keyword).  Table~\ref{tab:extname} lists the
different \Fits extensions (actually \Fits binary tables) used to implement
the \Origin input file format.  The revision number of the \Origin input
file format described in this document is \texttt{EXTVER=1} and is
indicated by the value of \Fits keyword \texttt{EXTVER} in each HDU
extension. The various extensions used in \Origin input data files are
detailed in the subsequent subsections.

\begin{table*}
\caption[Primary HDU header]{Example of the header part of a primary HDU in
  an \Origin input file.}
\label{tab:primary-hdu}
\renewcommand{\footnoterule}{}  
\begin{SmallText}
\begin{verbatim}
SIMPLE  =                    T / true FITS file (http://fits.gsfc.nasa.gov/)
BITPIX  =                    8 / 8-bit twos complement binary unsigned integer
NAXIS   =                    0 / this HDU contains no data
EXTEND  =                    T / this file may contain FITS extensions
BLANK
COMMENT This is an input FITS file for ORIGIN software for exoplanetary
COMMENT system sky energy distribution modeling.
BLANK
HISTORY Created by SOMEBODY on SOMEDATE.
END
\end{verbatim}
\end{SmallText}
\end{table*}

\begin{table}
  \centering
  \caption[\Fits extensions of the \Origin input file format]{Description
    of \Fits extensions of \Origin input file format.  All these extensions
    are saved into a \Fits binary table identified by its name, which is the
    value of the keyword \texttt{EXTNAME}. Column \textit{Number} indicates
    the number of extensions of a given type allowed in the
    file.\label{tab:extname}}
  \begin{tabular}{lcl}
  \hline\hline
  Extension Name & Number & Description\\
  \hline
  \texttt{'SCENARIO'}   & 1 & global parameters \\
  \texttt{'STAR'}       & 1 & star parameters \\
  \texttt{'EXO-ZODI'}   & 1 & exozodiacal cloud parameters \\
  \texttt{'LOCL-ZODI'}  & 1 & local zodiacal cloud parameters \\
  \texttt{'UNRESOLVED'} & any & parameters for planets \\
                              & or other point-\\
                              & like objects\\
  \hline
  \end{tabular}
\end{table}

\subsection{Scenario Table}
\label{s:scenario}

The parameters defining the scenario of an exoplanetary system observation
are stored into a \Fits extension named \texttt{'SCENARIO'}.  The
corresponding binary table contains a first column with the central
wavelength of the channels, and a second with their width. A number of scalar
parameters are also provided in the header part of this HDU:

\begin{itemize}
\item The value of \texttt{L2-FLAG} specifies whether the observer's
  position is at the L2 point (see Sect.~\ref{s:coordsyst}). Otherwise, Earth
  position is assumed.

\item The value of \texttt{DW-EPOCH} is the Julian date of the beginning of
  the operational phase of the instrument, as an epoch that later mission
  events will be relative to (DW comes from Darwin, the mission for which
  this standard was initially developed).

\item The value of \texttt{OBS-DATE} is the time (in fractional JD) from
  \texttt{DW-EPOCH} when the current observation starts. \Origin uses
  compact computer algorithms by \citet{Fliegel:vanFlandern:1968} for
  converting between Julian days and Gregorian calendar dates.

\item The value of \texttt{OBS-STEP} is the duration between two successive
  snapshot outputs of the scene, for following orbital motion of planets.

\item \texttt{STR-RES} is the resolution at which the flux calculator (as
  the \Origin software) generates a chromatic sky energy distribution (an
  image cube) of the star.

\item \texttt{IMG-RES} and \texttt{IMG-FOV} are the resolution
  and FOV, respectively, of the image (cube) of the exozodiacal dust. Also,
  \texttt{IMG-FOV} is used to compute the FOVs solid angle for uniform
  contribution calculation (see Sect.~\ref{s:outputlz}).

\item \texttt{CLOUD-DZ} is the integration step along the line of sight
  through 3-D dust distributions.

\end{itemize}

These parameters are listed in Table~\ref{tab:scenario-keywords}. Table~\ref{tab:scenario-configuration-header} shows a typical \texttt{SCENARIO}
header of an \Origin input file.  Note that, in this header, the value of
\texttt{NAXIS2} is also the number of effective spectral channels for which
output will be generated.  The columns of the binary table in a
\texttt{SCENARIO} extension are listed in
Table~\ref{tab:scenario-configuration-columns}.

\begin{table}
\caption[Scenario configuration Keywords]{List of \Fits keywords used to
    define scalar parameters of the \texttt{'SCENARIO'} extension.
    \label{tab:scenario-keywords}} \centering
  \begin{tabular}{lll}
  \hline\hline
  Keyword & Description & Units \\ 
  \hline
  \texttt{L2-FLAG}   & observer at L2 point & \\ 
  \texttt{DW-EPOCH}  & date of beginning of mission & JD \\ 
  \texttt{OBS-DATE}  & observation date (from \texttt{DW-EPOCH}) & JD \\ 
  \texttt{OBS-STEP}  & time between 2 snapshot outputs & hrs \\ 
  \texttt{OBS-NB}    & number of snapshot outputs &  \\ 
  \texttt{STR-RES}   & star resolution & \arcsec \\
  \texttt{IMG-RES}   & resolution of the zodiacal image & \arcsec \\
  \texttt{IMG-FOV}   & FOV & \arcsec \\
  \texttt{CLOUD-DZ}  & z integration step through cloud & AU \\
  \hline
  \end{tabular}
\end{table}

\begin{table*}
\caption[Scenario configuration header]{Typical \texttt{SCENARIO}
  header in an \Origin input file.}
\label{tab:scenario-configuration-header}
\centering
\renewcommand{\footnoterule}{}  
\begin{SmallText}
\begin{verbatim}
XTENSION= 'BINTABLE'           / FITS 3D BINARY TABLE
BITPIX  =                    8 / Binary data
NAXIS   =                    2 / Table is a matrix
NAXIS1  =                   16 / Width of table in bytes
NAXIS2  =                   13 / Number of entries in table
PCOUNT  =                    0 / Random parameter count
GCOUNT  =                    1 / Group count
TFIELDS =                    2 / Number of fields in each row
EXTNAME = 'SCENARIO'           / Table name
EXTVER  =                    1 / Version number of table
TFORM1  = 'D       '           / Data type for field
TTYPE1  = 'SPCH-CWL'           / Label for field
TUNIT1  = 'm       '           / Physical units for field
TFORM2  = 'D       '           /  Data type for field
TTYPE2  = 'SPCH-WDT'           / Label for field
TUNIT2  = 'm       '           / Physical units for field
L2-FLAG = 'F       '           / Observer at L2 point?
DW-EPOCH=              2456000 / Beginning of mission, JD
OBS-DATE=                  100 / Beginning of observation, from DW-EPOCH
OBS-STEP=                   24 /
OBS-NB  =                    1
STR-RES =                 8E-6
IMG-FOV =                0.715
IMG-RES =                 0.01
ORB-FOV =                    2
ORB-RES =                 0.01
CLOUD-DZ=                  0.1 / added AU
END
\end{verbatim}
\end{SmallText}
\end{table*}

\begin{table}
\caption[Scenario configuration columns]{Description of \Fits
    binary table for the \textsc{scenario} extension of the \Origin
    input file format.}
\label{tab:scenario-configuration-columns}
  \centering
  \begin{tabular}{llll}
  \hline\hline
  Column & Description &  Units \\
  \hline
  \texttt{SPCH-CWL} & central wavelength of channels & m \\
  \texttt{SPCH-WDT} & bandwidths of channels & m \\
  \hline
  \end{tabular}
\end{table}

\subsection{Star Parameters}
\label{s:param_star}

The star model parameters are stored in a \texttt{'STAR'} extension.
Table~\ref{tab:star-params} lists the keywords of the star model.

\begin{table*}
\caption{Star model keywords.}
\label{tab:star-params}
  \begin{center}
    \small
    \begin{tabular}{lll}
      \hline
      symbol & description & units \\
      \hline
      \texttt{ICRS- ALP, BET} & ICRS star coordinates & deg \\
      \texttt{SHIFT- ALP, BET} & offset of the star image from the pointing direction of the instrument & deg \\
      \texttt{SP-MODE} & whether black body or provided spectrum should be used & flag \\
      \texttt{LIMB-DRK} & which (uniform, linear, quadratic, etc.) limb darkening law should be used & flag \\
      \texttt{RLT-FLAG} & which among luminosity, star radius or effective temperature should be calculated from the others & flag \\
      \texttt{MGR} & which among mass, surface gravity or radius should be calculated from the others & flag \\
      \texttt{ML} & which from mass or luminosity  should be calculated from the other & flag \\
      \texttt{DISTANCE} & distance from observer & pc\\
      \texttt{EFF-TEMP} & star effective temperature & $\dgr\mathrm{K}$\\
      \texttt{MASS}     & mass of the star & $\Msun$ \\
      \texttt{LUMINOSI} & star luminosity & $\Lsun$ \\
      \texttt{RADIUS}   & star radius & $\Rsun$\\
      \texttt{LOG-GRAV} & log of star surface gravity & $g_\odot$\\
      \hline
    \end{tabular}
  \end{center}
\end{table*}

\Origin provides a database of monochromatic limb-darkening parameters and
specific intensity for stars of various spectral type and luminous class
(see Sect.~\ref{s:limb-darkening}).  This database is built from
\citet{VanHamme} tables and from a model of the HR diagram of existing
stars, to establish the relation between spectral type and luminosity class,
and physical parameters such as the star's effective temperature, surface
gravity and luminosity. Tables~\ref{t:mean-flux} displays, for reference,
the mean specific intensity in photometric bands, for the spectral types of
stars envisioned for exoplanet search.

\begin{table*}
  \renewcommand{\SIZE}[1]{{\tiny #1}}
  \renewcommand{\MC}[3]{\multicolumn{#1}{#2}{\SIZE{#3}}}
  \renewcommand{\X}[2]{\SIZE{$#1\cdot10^{#2}$}}
  \caption{Mean specific intensity in photometric bands (units: \New{$\mathrm{ph}\,\mathrm{s}^{-1}\,\mathrm{m}^{-2}\,\mathrm{sr}^{-1}\,\micron^{-1}$}).}
  \label{t:mean-flux}
  \centering
  \begin{tabular}{rllllllllllll}
    \hline\hline
    \MC{1}{c}{Type} &
    \MC{1}{c}{U} &
    \MC{1}{c}{B} &
    \MC{1}{c}{V} &
    \MC{1}{c}{R} &
    \MC{1}{c}{I} &
    \MC{1}{c}{J} &
    \MC{1}{c}{H} &
    \MC{1}{c}{K} &
    \MC{1}{c}{L} &
    \MC{1}{c}{M} &
    \MC{1}{c}{N} &
    \MC{1}{c}{Q} \\
    \hline
    \SIZE{F0V} & \X{1.5}{26} & \X{2.6}{26} & \X{2.4}{26} & \X{1.8}{26}
               & \X{1.2}{26} & \X{7.1}{25} & \X{3.9}{25} & \X{1.9}{25}
               & \X{5.5}{24} & \X{1.8}{24} & \X{2.4}{23} & \X{3.9}{22}\\
    \SIZE{F2V} & \X{1.3}{26} & \X{2.2}{26} & \X{2.1}{26} & \X{1.6}{26}
               & \X{1.1}{26} & \X{6.7}{25} & \X{3.7}{25} & \X{1.8}{25}
               & \X{5.4}{24} & \X{1.8}{24} & \X{2.3}{23} & \X{3.7}{22}\\
    \SIZE{F5V} & \X{8.6}{25} & \X{1.5}{26} & \X{1.6}{26} & \X{1.3}{26}
               & \X{9.6}{25} & \X{5.8}{25} & \X{3.4}{25} & \X{1.7}{25}
               & \X{5.0}{24} & \X{1.6}{24} & \X{2.1}{23} & \X{3.5}{22}\\
    \SIZE{F8V} & \X{6.7}{25} & \X{1.2}{26} & \X{1.3}{26} & \X{1.1}{26}
               & \X{8.7}{25} & \X{5.4}{25} & \X{3.3}{25} & \X{1.6}{25}
               & \X{4.8}{24} & \X{1.5}{24} & \X{2.0}{23} & \X{3.3}{22}\\
    \SIZE{G0V} & \X{5.1}{25} & \X{9.6}{25} & \X{1.1}{26} & \X{1.0}{26}
               & \X{7.8}{25} & \X{5.0}{25} & \X{3.1}{25} & \X{1.5}{25}
               & \X{4.6}{24} & \X{1.5}{24} & \X{2.0}{23} & \X{3.2}{22}\\
    \SIZE{G2V} & \X{3.7}{25} & \X{7.5}{25} & \X{9.3}{25} & \X{8.6}{25}
               & \X{6.9}{25} & \X{4.6}{25} & \X{3.0}{25} & \X{1.4}{25}
               & \X{4.4}{24} & \X{1.4}{24} & \X{1.9}{23} & \X{3.1}{22}\\
    \SIZE{G5V} & \X{3.7}{25} & \X{7.5}{25} & \X{9.3}{25} & \X{8.6}{25}
               & \X{6.9}{25} & \X{4.6}{25} & \X{3.0}{25} & \X{1.4}{25}
               & \X{4.4}{24} & \X{1.4}{24} & \X{1.9}{23} & \X{3.1}{22}\\
    \SIZE{G8V} & \X{2.5}{25} & \X{5.7}{25} & \X{7.6}{25} & \X{7.4}{25}
               & \X{6.0}{25} & \X{4.2}{25} & \X{2.8}{25} & \X{1.4}{25}
               & \X{4.2}{24} & \X{1.3}{24} & \X{1.8}{23} & \X{2.9}{22}\\
    \SIZE{K0V} & \X{1.7}{25} & \X{4.3}{25} & \X{6.0}{25} & \X{6.1}{25}
               & \X{5.2}{25} & \X{3.8}{25} & \X{2.6}{25} & \X{1.3}{25}
               & \X{4.0}{24} & \X{1.2}{24} & \X{1.7}{23} & \X{2.8}{22}\\
    \SIZE{K1V} & \X{1.0}{25} & \X{3.1}{25} & \X{4.7}{25} & \X{5.0}{25}
               & \X{4.4}{25} & \X{3.4}{25} & \X{2.5}{25} & \X{1.2}{25}
               & \X{3.8}{24} & \X{1.1}{24} & \X{1.6}{23} & \X{2.7}{22}\\
    \SIZE{K2V} & \X{1.0}{25} & \X{3.1}{25} & \X{4.7}{25} & \X{5.0}{25}
               & \X{4.4}{25} & \X{3.4}{25} & \X{2.5}{25} & \X{1.2}{25}
               & \X{3.8}{24} & \X{1.1}{24} & \X{1.6}{23} & \X{2.7}{22}\\
    \SIZE{K3V} & \X{6.0}{24} & \X{2.1}{25} & \X{3.4}{25} & \X{4.0}{25}
               & \X{3.7}{25} & \X{3.0}{25} & \X{2.3}{25} & \X{1.2}{25}
               & \X{3.6}{24} & \X{1.1}{24} & \X{1.5}{23} & \X{2.5}{22}\\
    \SIZE{K4V} & \X{3.2}{24} & \X{1.4}{25} & \X{2.4}{25} & \X{3.1}{25}
               & \X{3.0}{25} & \X{2.6}{25} & \X{2.2}{25} & \X{1.1}{25}
               & \X{3.4}{24} & \X{9.9}{23} & \X{1.5}{23} & \X{2.4}{22}\\
    \SIZE{K5V} & \X{1.6}{24} & \X{8.1}{24} & \X{1.6}{25} & \X{2.2}{25}
               & \X{2.4}{25} & \X{2.2}{25} & \X{2.0}{25} & \X{1.0}{25}
               & \X{3.2}{24} & \X{9.2}{23} & \X{1.4}{23} & \X{2.3}{22}\\
    \SIZE{K7V} & \X{7.6}{23} & \X{4.4}{24} & \X{9.9}{24} & \X{1.5}{25}
               & \X{1.8}{25} & \X{1.8}{25} & \X{1.7}{25} & \X{8.8}{24}
               & \X{2.9}{24} & \X{8.6}{23} & \X{1.3}{23} & \X{2.2}{22}\\
    \SIZE{M0V} & \X{3.9}{23} & \X{2.3}{24} & \X{5.8}{24} & \X{9.6}{24}
               & \X{1.3}{25} & \X{1.4}{25} & \X{1.3}{25} & \X{7.4}{24}
               & \X{2.6}{24} & \X{8.0}{23} & \X{1.3}{23} & \X{2.1}{22}\\
    \SIZE{M1V} & \X{3.9}{23} & \X{2.3}{24} & \X{5.8}{24} & \X{9.6}{24}
               & \X{1.3}{25} & \X{1.4}{25} & \X{1.3}{25} & \X{7.4}{24}
               & \X{2.6}{24} & \X{8.0}{23} & \X{1.3}{23} & \X{2.1}{22}\\
    \SIZE{M2V} & \X{1.8}{23} & \X{1.2}{24} & \X{3.2}{24} & \X{5.5}{24}
               & \X{9.4}{24} & \X{1.1}{25} & \X{1.1}{25} & \X{6.2}{24}
               & \X{2.3}{24} & \X{7.3}{23} & \X{1.2}{23} & \X{2.0}{22}\\
    \SIZE{M3V} & \X{1.8}{23} & \X{1.2}{24} & \X{3.2}{24} & \X{5.5}{24}
               & \X{9.4}{24} & \X{1.1}{25} & \X{1.1}{25} & \X{6.2}{24}
               & \X{2.3}{24} & \X{7.3}{23} & \X{1.2}{23} & \X{2.0}{22}\\
    \SIZE{M4V} & \X{1.7}{23} & \X{1.1}{24} & \X{3.2}{24} & \X{5.7}{24}
               & \X{9.5}{24} & \X{1.1}{25} & \X{1.0}{25} & \X{6.1}{24}
               & \X{2.3}{24} & \X{7.4}{23} & \X{1.2}{23} & \X{2.0}{22}\\
    \hline
  \end{tabular}
\end{table*}

The star model can be built by choosing one of the items from the stellar
database provided with the \Origin software, or by specifying the
star parameters (either in the same format as the database or, more simply,
by a simple black body emission model characterized by the star effective
temperature, luminosity or radius, mass or surface gravity and, optionally,
limb-darkening parameters).

\subsection{(Exo)zodiacal Cloud}

Exo- and local zodiacal cloud parameters are stored into
\texttt{EXTNAME='EXO-ZODI'} and \texttt{EXTNAME='LOCAL-ZODI'} extensions.
Table~\ref{t:zodi-key} lists the keywords of the (exo)zodiacal model. Most
parameters are those of the implemented \citet{Kelsall} model.  The binary
table of these extensions list solar system's \citet{Kelsall} parameters of the three dust bands of the  zodiacal cloud.

\begin{table*}
\caption{(Exo)zodiacal model keywords.}
\label{t:zodi-key}
  \begin{center}
    \small
    \begin{tabular}{cll}
      \hline
      symbol & description & units \\
      \hline
      \texttt{TYPE} & whether this extension is a local or exo- zodiacal cloud & flag \\
      \texttt{ZODI} & mean density of dust & zodi \\
      \texttt{DUST-ST} & dust sublimation temperature & $\dgr\mathrm{K}$ \\
      \texttt{DUST-RT} & dust reference temperature & $\dgr\mathrm{K}$ \\
      \texttt{ETL} & exponent temperature law coefficient & unitless \\
      \texttt{CLOUD-O} & cloud outer radius & AU\\
      \texttt{SC-*}    & smooth cloud parameters of the Kelsall model, not all listed here & \\
      \hline
    \end{tabular}
  \end{center}
\end{table*}

\subsection{Unresolved Sources}

Extension(s) \texttt{EXTNAME='UNRESOLVED'} store the parameters for
unresolved sources such as planets. The format allows the specification of all
their orbital parameters, but can also be used to account for fixed
background point-like sources such as field stars or distant galaxies.
Table~\ref{t:unres-key} lists the keywords of the unresolved sources model.
In addition to these, \citet{Kelsall} parameters of the circumsolar dust
ring and blob can be specified.  \Origin uses the \citet{Kelsall} model
equations, except the radius of the ring is the semi-major axis of the
planet.  Table~\ref{t:unres-bintab} lists the columns of the binary table
of the unresolved source extension: this is the optionally provided planetary
spectrum.

\begin{table*}
\caption{Unresolved sources (planets) model keywords.}
\label{t:unres-key}
  \begin{center}
    \small
    \begin{tabular}{lll}
      \hline
      symbol & description & units \\
      \hline
      \texttt{ORB-APER} & which, from semi-major axis or period is computed from the other & flag \\
      \texttt{P-RADIUS} & planet radius & $R_{\oplus}$ \\
      \texttt{P-TEMP}   & planet temperature & $\dgr\mathrm{K}$ \\
      \texttt{P-ALBEDO} & planet albedo & unitless \\
      \texttt{ORB-T0}   & epoch of periastron passage & Julian date \\
      \texttt{ORB-OMA}  & position angle of the ascending node & deg \\
      \texttt{ORB-OMP}  & argument of periastron & deg \\
      \texttt{ORB-I}    & inclination of orbit & deg \\
      \texttt{ORB-E}    & eccentricity & unitless \\
      \texttt{ORB-A}    & semi-major axis & AU \\
      \texttt{ORB-PER}  & orbital period & days  \\
      \texttt{SP-FLAG}  & whether to use specific intensities provided in the extensions binary table, or black body emission & flag \\
      \hline
    \end{tabular}
  \end{center}
\end{table*}

\begin{table}
  \caption[Provided planetary spectrum definition]{Description of \Fits binary
    table for \textsc{unresolved} extension in the \Origin input file format.}
  \label{t:unres-bintab}
  \centering
  \begin{tabular}{llll}
  \hline\hline
  Column & Description &  Units \\
  \hline
  \texttt{LAMBDA}  & channel's central wavelength & m \\
  \texttt{SP\_FLUX} & specific intensity & ph s$^{-1}$ m$^{-2}$ sr$^{-1}$ \micron$^{-1}$\\
  \hline
  \end{tabular}
\end{table}

\section{Output Format}
\label{s:output}

The output format of the \Origin software is also proposed as an input
standard for instrumental simulators.  It consists of two \Fits files (or
two series of files, if several observations at different times are
demanded).  The \texttt{ORBIT} file simply contains an image of the planets'
orbits with the current position of each planet, and it mainly serves a
human visual check purpose. The \texttt{LAYERED} file contains the physical
information that an instrumental simulator needs, and is described in the
following.

We first note that the dynamic in specific intensity and resolution between
the different sources in a planetary system scene is such that it is
impractical to compute a global image, with a given resolution, of the
whole FOV, and submit it to an instrumental simulator. Following the input
standard, the output is also layered, as a \Fits file containing three types of
descriptions of incoming fluxes:
\emph{resolved sources}, \emph{unresolved sources}, and \emph{FOV-uniform}
contributions, that can all be used by an instrumental simulator.

These files, being generated by the \Origin software, are quite
explicit, so this section is considerably more straightforward than the
previous.

\subsection{Primary HDU}
\label{s:primary}

Unlike the input format, the primary HDU restates some of the data from the
input format. Table~\ref{t:primary} gives an example of such a header.

\begin{table*}
\caption[Primary output format header]{Typical primary header in
  an \Origin output file.}
\label{t:primary}
\centering
\renewcommand{\footnoterule}{}  
\begin{SmallText}
\begin{verbatim}
SIMPLE  =                    T / Java FITS: Mon May 22 07:38:11 CEST 2006
BITPIX  =                    8
NAXIS   =                    0 / Dimensionality
EXTEND  =                    T / file contains FITS extensions
COMMENT = 'ORIGIN LAYERED OUTPUT FILE' /
ORIGIN  = 'ORIGIN v1.0'        / Name and version of software
AUTHOR  = 'Author  '           /
DATE    = 'Mon May 22 07:38:10 CEST 2006' / File creation date
CTRLFILE= 'C:\ORIGIN\Livraison\Full Scenarii\Gl_876.fits' / Name and path of ORI
DW-EPOCH=            2456000.0 / Darwin epoch [JD]
OBS-DATE=                100.0 / Observation date since epoch [days]
IMG-NBR =                    6 / Number of images generated during simulation
IMG-RANK=                    1 / Rank of current image simulation
END
\end{verbatim}
\end{SmallText}
\end{table*}

\subsection{Output Spectral Channels}

In the output file, the \texttt{EXTNAME='SPCHANNELS'} extension is a binary
table which gives the effective central wavelengths and spectral
bandwidths of the simulated model.  This information is similar to that specified in the input format (see Sect.~\ref{s:scenario}), and is not
further described here.

\subsection{Star or Exo-zodiacal Output Maps}

In the output file, the \texttt{EXTNAME='RESOLVED~OBJECT~LAYER'} extensions
are \Fits \textit{images} containing 3-D maps (right ascension, declination
and wavelength) of the star's limb darkened photosphere, or of the
exozodiacal dust cloud.  Table~\ref{tab:resolved} shows an example of the
header of this extension.  The number of pixels in the image is always odd,
with the \emph{reference pixel} marking the center of the image.  For an
image example, refer to Fig.~\ref{f:origin_star}.

\begin{table*}
\caption[Resolved configuration header]{Typical \texttt{'RESOLVED~OBJECT~LAYER'}
  header in an \Origin output file.}
\label{tab:resolved}
\centering
\renewcommand{\footnoterule}{}  
\begin{SmallText}
\begin{verbatim}
XTENSION= 'IMAGE   '           / Java FITS: Wed Mar 15 10:35:36 CET 2006
BITPIX  =                  -64
NAXIS   =                    3 / Dimensionality
NAXIS1  =                  121
NAXIS2  =                  121
NAXIS3  =                   13
PCOUNT  =                    0 / No extra parameters
GCOUNT  =                    1 / One group
EXTNAME = 'RESOLVED OBJECT LAYER'/This Layer contains the star image
EXTVER  =                    1 /
CTYPE1  = 'RA      '           / Right ascension axis
CRPIX1  =                   61 / Reference pixel along axis 1 (starting from 1)
CRVAL1  =                 75.0 / Coordinate of ref. pixel along axis 1 [deg]
CDELT1  = -2.222222222222222E-9/ Pixel step along axis 1 [deg]
CTYPE2  = 'DEC     '           / Declination axis
CRPIX2  =                   61 / Reference pixel along axis 2 (starting from 1)
CRVAL2  =                155.0 / Coordinate of ref. pixel along axis 2 [deg]
CDELT2  = 2.222222222222222E-9 / Pixel step along axis 2 [deg]
END
\end{verbatim}
\end{SmallText}
\end{table*}

\subsection{Local Zodiacal and Galactic Background}
\label{s:outputlz}

The background emission of the local zodiacal light, which is uniform over
the field of view, is provided by the output format in
$\mathrm{ph}\,\mathrm{s}^{-1}\,\mathrm{m}^{-2}$ per spectral channel in the
binary table extension \texttt{EXTNAME='CONSTANT~LAYER'}
(Fig.~\ref{f:local_result}).  For that, the \Origin software considers the
FOV solid angle defined in the \texttt{EXOZODI} extension in the input
format (see Sect.~\ref{s:scenario}).  The header of this extension contains only the binary tables column definitions (Table~\ref{t:constant_col}).

\begin{table}
  \caption[Constant layer columns]{Description of \Fits binary
    table columns for \textsc{constant layer} extension in the \Origin output file format.}
  \label{t:constant_col}
  \centering
  \begin{tabular}{llll}
  \hline\hline
  Column & Description &  Units \\
  \hline
  \texttt{SPCHCWL}  & channel's central wavelength & m \\
  \texttt{SPCHWDT}  & channel's width              & m \\
  \texttt{SPCHFLUX} & flux                         & ph s$^{-1}$ m$^{-2}$\\
  \hline
  \end{tabular}
\end{table}

The galactic background can be manually inserted here using this same extension format, since it requires no calculation by the \Origin software.

\subsection{Planets}

The output file format contains as many \texttt{UNRESOLVED~OBJECT~LAYER} binary table extensions as there were \texttt{UNRESOLVED} planet definition extensions in the corresponding input file.
Table~\ref{tab:unresolved} shows the header of such a planet binary table output extension. It provides the precise
coordinates in the field of view of the unresolved source (fields
\texttt{RA-STR} and \texttt{DEC-STR}).  The binary table itself contains the
calculated flux for the planet, degraded to the resolution indicated by the
primary extension (Sect.~\ref{s:primary}). It can be seen that the columns of this table are exactly the same as those of the \texttt{CONSTANT~LAYER} extension (Sect.~\ref{s:outputlz} and Table~\ref{t:constant_col}).

\begin{table*}
\caption[Unresolved configuration header]{Typical \texttt{'UNRESOLVED'}
  header in an \Origin output file.}
\label{tab:unresolved}
\centering
\renewcommand{\footnoterule}{}  
\begin{SmallText}
\begin{verbatim}
XTENSION= 'BINTABLE'           / Java FITS: Wed Mar 15 10:35:43 CET 2006
BITPIX  =                    8
NAXIS   =                    2 / Dimensionality
NAXIS1  =                   24
NAXIS2  =                   13
PCOUNT  =                    0
GCOUNT  =                    1
TFIELDS =                    3
TFORM1  = '1D      '
TDIM1   = '(1)     '
TFORM2  = '1D      '
TDIM2   = '(1)     '
TFORM3  = '1D      '
TDIM3   = '(1)     '
EXTNAME = 'UNRESOLVED OBJECT LAYER'/This Layer contains the spectrum of a planet
EXTVER  =                    1 /
RA      =    75.00005875513867 / Object Absolute Right Ascension [deg]
DEC     =    155.0001242618757 / Object Absolute Declination [deg]
RA-STR  =  0.21151849913006165 / Object Right Ascension from Star [Arcsec]
DEC-STR =   0.4473427524966594 / Object Declination from Star [Arcsec]
TTYPE1  = 'SPCHCWL '           / Spectral channel central wavelength
TUNIT1  = 'm       '           /
TTYPE2  = 'SPCHWDT '           / Spectral channel width
TUNIT2  = 'm       '           /
TTYPE3  = 'SPFLUX  '           / Object Flux
TUNIT3  = 'count/m^2/s'        /
END
\end{verbatim}
\end{SmallText}
\end{table*}

\end{appendix}

\end{document}